\documentclass[aps,prb,preprint,onecolumn,superscriptaddress]{revtex4}
\usepackage{graphicx}
\usepackage{units}
\usepackage{color}
\usepackage{amsmath}

\begin{document}

\title{Modeling the Dynamics of Trapped electrons in Quantum Dots}

\author{R. Carmina Monreal}
\email[Corresponding Author: ]{r.c.monreal@uam.es}

\affiliation{Departamento de F\'{\i}sica Te\'{o}rica de la Materia Condensada and Condensed Matter Physics Center (IFIMAC), Universidad Aut\'{o}noma de Madrid, E-28049 Madrid, Spain}

\begin{abstract}
We analyze the effects of electron-electron and electron-phonon interactions in the dynamics of a system of two or three electrons that can be trapped to a localized state and detrapped to extended band states of a quantum dot using a simple model in which the valence and conduction bands of the quantum dot are modeled by two single-particle energy levels and the trap is described by one single-particle level within the band gap. 
In spite of its simplicity the time dependent problem has no analytical solution but a numerically exact one can be found at a relatively low computational cost. Within this model, we study the time evolution of the electron occupancies of conduction and valence bands and the trap state, as well as the statistical factors influencing light emission of different energies. 
In most of the analyzed cases, the system dynamics has a very short transient determined by the hopping parameters, that can be of tens of femtoseconds,
followed by a quasi-stationary regime in which the electron occupancies either oscillate periodically around their time-averaged values or remain nearly constant. 
We find signatures of strong electronic correlations in the electronic motion for negative values of the effective electron-electron Coulomb interaction that are not translated to the statistical factors for light emission. These factors always show fast oscillations associated to electrons hopping back and forth from the conduction band to the trap, irrespective of the motion of the valence band electrons. 
Our calculations show that light emission of different energies is always possible except in the especial cases in which the valence band is initially filled with two electrons. In these cases the valence band can lose and recover electrons periodically but exciton emission is negligible at any time. We use this fact to attempt to give a possible explanation for the increase in the intensity of exciton emission with the concomitant decrease in the intensity of the green emission lines upon continuous illumination with ultraviolet radiation, 
experimentally observed for ZnO nanoparticles suspended in an alcohol.

\end{abstract}

\maketitle

\newpage

\section{Introduction}

The electronic and optical properties of colloidal semiconductor nanocrystals (quantum dots QDs)\cite{Pichaardy, LEED3} are attracting an increasing interest for applications in optoelectronic devices\cite{Opto1,Opto2}, solar cells \cite{Solar1} and light emitting diodes \cite{LEED1,LEED2} among others.
 Light emission in colloidal QDs is conditioned by the presence of defects and/or impurities at the surface. 
In these small systems, the large surface to volume ratio results in relative abundance of surface defects of varied microscopic nature, such as lattice relaxations, impurity atoms or surface ligands, generically called "traps", that produce states in the band gap \cite{LEED3, JPCB_104_1715_vanDijken}.
Once the QD is excited by light across the band gap, these localized states can trap charge carriers and alter the subsequent light emission. 
 A paradigmatic case is ZnO
 for which its luminescent properties are known to depend upon the atmosphere surrounding the particles
\cite{ JPCB_104_1715_vanDijken, ChemPhysLett_122_507_koch, JPhysChem_91_3789_bahnemann, JLumin_90_123_vanDijken, JPCB_104_4355_vanDijken}, 
 the charge on them \cite{JPCC_114_220_stroyuk, JPCC_115_21635_yamamoto,
JPCC_116_20633_cohn}, the passivation of surface defects \cite{JPCB_109_20810_norberg} or chemical modifications at the surface \cite{JACS_123_11651_shim}.
In ZnO two well separated emission bands appear: the narrow band (exciton band) is in the ultraviolet and 
originates from radiative recombination of excited electrons with the holes left in the valence band, while the usually prominent broad band 
is in the visible (green band) and arises 
from recombination of excited electrons with deep-trapped holes whose electronic states lie in the middle of the band gap and are therefore localized. 
The relative intensity of both bands depends
on the above mentioned environmental conditions and  on the illumination time \cite{JPCC_114_220_stroyuk, JPCC_115_21635_yamamoto,
JPCC_116_20633_cohn}. 
Photoexcitation of colloidal ZnO QDs under continuous illumination with ultraviolet radiation yields the quenching of the prominent green band 
while the excitonic band increases in intensity. Furthermore an infrared absorption band develops due to the accumulation of electrons in the conduction band,
that forms an ultra low-density electron gas \cite{JPCL_4_3024_faucheux, JPCL_5_976_faucheaux, ACSNano_8_1065_schimpf}. 
An interesting experimental finding is that the density of this electron gas depends on the reductant agent added to the colloidal nanocrystals, usually an alcohol, which scavenges the produced holes irreversibly in times $\lesssim 15$ ps for EtOH \cite{JACS_Schimpf}.
Therefore, a key point to understand the light emission properties of QDs is to know the timescales and efficiencies of possible mechanisms competing with radiative decay in the filling of the states because if these mechanisms were much faster the photoluminescence will be quenched. 
One of the proposed mechanisms is a trap-assisted Auger recombination process in which a conduction band electron fills the deep-trap hole with the released energy transferred to another conduction band electron \cite{JPCC_116_20633_cohn, Nanolett_cohn}, with an estimated time of $\simeq 150$ ps for ZnO nanocrystals in EtOH \cite{JPCC_115_21635_yamamoto}.
This value is on the order of magnitude as for other Auger recombination processes of electrons and holes in semiconductor nanocrystals. Auger times for trions in CdSe range between 5 ps and 1000 ps \cite{Nanolett_cohn_1, Nanolett_Vaxemburg}, 
while the times for Auger recombination of biexcitons  are from 5 ps to 500 ps for nanocrystals of 1 nm to 3 nm in size \cite{ChemRev_Guyot, Nanolett_Rabani, Science_Klimov, PRL_Htoon,PRB_Taguchi,Zhu}. 
Another mechanism has been explored in a recent calculation by du Foss\'e at al.  \cite{ChemMat_duFosse}, where the initial photoexcitation of an electron to the conduction band of a CdSe QD of ca. 2 nm in diameter leads to a transient trap state localized in a Cd-Cd dimer that appears and disappears on the picoseconds time scale. 
However, the existence of local lattice relaxations at the trap site strongly recalls electron-phonon interaction. Although the vibronic interaction has already been included in first-principles calculations of photoluminescence line shapes of a number of defects in bulk semiconductors  \cite{Janotti_PRL, Janotti_APL,Janotti_JAP}, all of these studies have in common that they focus on the static electronic or electronic plus vibronic configurations of the system. An equivalent analysis for a dynamical problem would be a formidable task nowadays even for a small system. Hence simple exactly solvable model Hamiltonians are useful to gain physical insight into the dynamics of electron-electron electron-phonon interactions. The aim of the present work is to propose a simple model Hamiltonian that can describe the main characteristics of the dynamics of electrons trapped to deep-traps in wide band gap semiconductor QDs using a minimum set of parameters.

In  previous works \cite{Monreal_PRB, Monreal_JPCC} we analyzed the effects of electron-phonon interaction in the dynamics of a system of few electrons that can be trapped and detrapped from the trap to the QD. In these works is was assumed that only the electrons excited to the conduction band upon photoexcitation were active, 
the holes left behind having been completely scavenged by the alcohol. However, the rising of the excitonic emission band under continuous illumination tells that not all the holes can be quenched. The purpose of the present work is to investigate the role that the holes in the valence band play in the system dynamics and in the luminescent properties of QDs using a simple model. 
The QD is described by two single-particle energy levels mimicking the valence and the conduction bands.
Substitution of a real QD by a few discrete energy levels has been frequent in the literature
\cite{Inoshita_PRB, Kral_PRB,Stauber_PRB, Vasilevskiy_PRB}. This description will be valid in situations where a few levels could be populated with appreciable probability in the initial photoexcitation event (as for example for photon energies slightly above the band gap) or by electronic hopping from the trap. 
The trap is modeled by one single-particle energy level within the band gap,
 as is the case for deep traps in semiconductor QDs
\cite{JPCB_104_1715_vanDijken, JLumin_90_123_vanDijken, JPCB_104_4355_vanDijken, JPCB_109_20810_norberg,Janotti_PRL,Janotti_APL}. Electrons in the trap are coupled to a single phonon mode with a coupling parameter $\lambda$, and
 QD and trap are connected by hopping parameters. This model does not consider any other agent present in an experimental set-up, that could transfer electrons or holes to/from the dot-trap system.
In the photoexcitation of a QD, electrons are excited to the conduction band. We take this event as our starting point and consider a number of electrons in our energy levels at time t=0. The electrons are subsequently allowed to hop back and forth to the trap and we calculate numerically the evolution of the coupled electron-phonon system as a function of time. The physical parameters of our problem are in principle unrestricted but we will use values appropriate for deep traps in wide band gap semiconductors QDs. Within this framework, we calculate the electron occupancies of the valence, conduction and trap levels. We also calculate 
the statistical factors influencing the intensity of the light emissions of the excitonic and the defect (or green) lines.
This minimal model allows us to investigate numerically both the short and the long time dynamics of the electrons, which will be presently infeasible for first-principles type of Hamiltonians. The article is organized in the following way. Section II expounds the theory. Our results are presented and discussed in Section III and our conclusions in Section IV. Appendixes A to E contain additional information not included in the main text.
Natural units $\hbar=e=m_e=k_{B}=1$ are used except otherwise indicated.

\section{Theoretical Methods}
\label{sec-theory}

The core of a QD of a few nanometers in size still presents the lattice structure of the bulk sample and its band structure shows Bloch characteristics even though the electronic states are quantizied due to size effects \cite{Efros_Rosen}.
In our model we substitute the full electronic structure of the QD by two single-particle levels of energies 
$\epsilon_{VB}$ and $\epsilon_{CB}$ representing the highest occupied "valence band" (VB) and the lowest unoccupied "conduction band" (CB) states for electrons, respectively ($\epsilon_{CB} > \epsilon_{VB}$), separated by the energy gap $E_g= \epsilon_{CB}- \epsilon_{VB}$. The deep trap
is described by one single-particle level of energy $\epsilon_T$ well within the band gap and is therefore localized. 
Electron-electron Coulomb repulsion $U$ between two electrons of opposite spins is considered in the trap localized state but it is neglected in the dot 
because of the Bloch nature of the electronic levels in real QDs. 
The interaction between an electron in the dot and another in the trap should be smaller than $U$ and it is also neglected. Note that we do not consider electron-electron interactions in the dot and thus exciton or biexciton binding energies are neglected.   Although several phonons could be active at the trap site, first-principles calculations of the photoluminescence line shapes of several defects in semiconductors \cite{Janotti_PRL,Janotti_APL,Janotti_JAP}, 
show that two phonons of similar energies and coupling constants are enough to fit the experiments. Then, in our model electrons in the trap couple to a single local phonon of energy
 $\omega_0$ with a coupling constant $\lambda$ \cite{Holstein}.
Electron-phonon interaction in the QD should be weaker than in the trap, where there are strong lattice distortions, and it is disregarded. Evidence of the so called phonon bottleneck effect (excited electrons are long lived excitations in a small QD as
they would couple very inefficiently to phonons due to the large mismatch between the inter-level energy spacing and the phonon energy \cite{Nozik_PhysE, Yang_NatPhot, arxiv}) also supports this assumption. 
 QD and trap are connected via hopping parameters $V_{CB}$ and $V_{VB}$, describing electron tunneling from/to the trap and the CB and the VB, respectively.
 Thus, the Hamiltonian has the form of the Anderson-Holstein impurity model and reads

\begin{equation}
\hat H=\hat H_{QD}+ \hat H_T+\hat H_{hop},
\label{Ham}
\end{equation}
where $\hat H_{QD}$ and $\hat H_T$  are the Hamiltonians of the uncoupled dot and trap subsystems respectively 

\begin{equation}
\hat H_{QD}=\epsilon_{VB} \sum_{\sigma}\hat n_{0 \sigma}+\epsilon_{CB} \sum_{\sigma}\hat n_{1 \sigma},
\label{Ham-D}
\end{equation}

\begin{equation}
\hat H_{T}= \epsilon_T \sum_{\sigma}\hat n_{T \sigma}+ U \hat n_{T \uparrow} \hat n_{T \downarrow}+
\omega_0 \hat b^{\dagger} \hat b+ \lambda \sum_{\sigma} \hat n_{T \sigma}(\hat b^{\dagger}+\hat b),
\label{Ham-T}
\end{equation}

and QD and trap are coupled by the hopping Hamiltonian $\hat H_{hop}$ 
\begin{equation}
\hat H_{hop}=V_{VB} \sum_{\sigma} (\hat c_{T \sigma}^{\dagger} \hat c_{0, \sigma} + hc.)+
V_{CB} \sum_{\sigma} (\hat c_{T \sigma}^{\dagger} \hat c_{1 \sigma} + hc.).
\label{Ham-hop}
\end{equation}

In Eqs. (\ref{Ham-D}), (\ref{Ham-T}) and (\ref{Ham-hop})  $\hat c_{0 \sigma}$ ($\hat c_{0 \sigma}^{\dagger}$), $\hat c_{1 \sigma}$ ($\hat c_{1 \sigma}^{\dagger}$) and 
$\hat c_{T}$ ($\hat c_{T}^{\dagger}$)
are the annihilation (creation) operators for electrons of spin $\sigma$ in the single-particle levels VB, CB and trap, respectively,
$\hat n_{0 \sigma}=\hat c_{0 \sigma}^{\dagger} \hat c_{0 \sigma}$, $\hat n_{1 \sigma}=\hat c_{1 \sigma}^{\dagger} \hat c_{1 \sigma}$ and 
$\hat n_{T \sigma}=\hat c_{T \sigma}^{\dagger} \hat c_{T \sigma}$ being their respective number operators,
$\hat b$ ($\hat b^{\dagger}$) are the annihilation (creator) operators for phonons, $\hat b^{\dagger} \hat b$ being their number operator.

For a better understanding of our results we quote here the so called atomic limit ($V_{CB}, V_{VB} \rightarrow 0$) in which an electron in a localized level interacts with phonons.
The stationary solution of Eq. (\ref{Ham-T}) \cite{Mahan, Hewson_Meyer} yields to renormalization of the electron energy to $\tilde \epsilon_T=\epsilon_T-\frac{\lambda^2}{\omega_0}$ 
and of the Coulomb repulsion $U$ to $U_{eff}=U-2\frac{\lambda^2}{\omega_0}$
and a local density of states which, at zero temperature, reads \cite{ Hewson_Meyer}

\begin{eqnarray}
\rho_{T \sigma}(\omega)=-\frac{1}{\pi} \lim_ {\eta \rightarrow 0} Im e^{-g} \sum_{n=0}^{\infty}  \frac{g^n}{n!} 
[ \frac{\langle(1-\hat n_{T \sigma})(1-\hat n_{T -\sigma}) \rangle}{\omega-\tilde \epsilon_T-n \omega_0+i\eta}
+\frac{\langle(1-\hat n_{T -\sigma})\hat n_{T \sigma} \rangle}{\omega-\tilde \epsilon_T+n \omega_0+i\eta} \nonumber \\
+\frac{\langle(1-\hat n_{T \sigma}) \hat n_{T -\sigma} \rangle}{\omega-\tilde \epsilon_T-U_{eff}-n \omega_0+i \eta}
+\frac{\langle\hat n_{T \sigma} \hat n_{T -\sigma} \rangle}{\omega-\tilde \epsilon_T-U_{eff}+n \omega_0+i\eta} ],
\label{rho}
\end{eqnarray}
where $g=(\frac{\lambda}{\omega_0})^2$ is the Huang-Rhys parameter and $Im$ stands for imaginary part.
The density of states consists of peaks (sub-levels) at
$\tilde \epsilon_T  \pm n \omega_0$ and  $\tilde \epsilon_T+U_{eff} \pm n \omega_0$ with weights distributed according to the Poisson distribution function but modified by the electron occupancy.
This structure will be seen in our results. The effect of temperature is to transfer spectral weight from the sub-levels with $n \simeq g$, where the density of states has a maximum at $T=0$, to the sub-levels with smaller and larger values of $n$  \cite{Mahan} . The effect depends exponentially on the ratio $\frac{\omega_0}{T}$ and is small  for  $\frac{\omega_0}{T} \ge 1$, which is the case at room temperature and below for the value of $\omega_0$ we will use in the calculations. Also, thermal excitation within the QD is small 
in our case because the gap energy, $E_{g}=\epsilon_{CB}-\epsilon_{VB}$, is much larger than $T$.
For these reasons and for the sake of simplicity, we will work a zero temperature.
We note that, while $U$ is a positive defined energy, $U_{eff}$ can be negative indicating an effective attractive 
interaction between electrons mediated by the electron-phonon interaction.

The time-dependent problem for Hamiltonian Eq. (\ref{Ham}) will be solved for a system of two and three electrons. For this small system a numerically exact solution can be
found for any value of the parameters by using a complete basis set containing all possible states of electrons and phonons. 
To this end, the time dependent wavefunction is written as a linear combination of these states and the time-dependent Schr\"odinger equation is projected onto each of them leading to a system of coupled linear differential equations for the combination coefficients that is solved numerically with given initial conditions.  
Several initial conditions will be imposed according to which of the single-particle states, with at least one electron in the CB, 
are occupied by the electrons upon the photoexcitation process, as detailed below.

\begin{figure}[htbp]
\centering
\includegraphics[width=100mm]{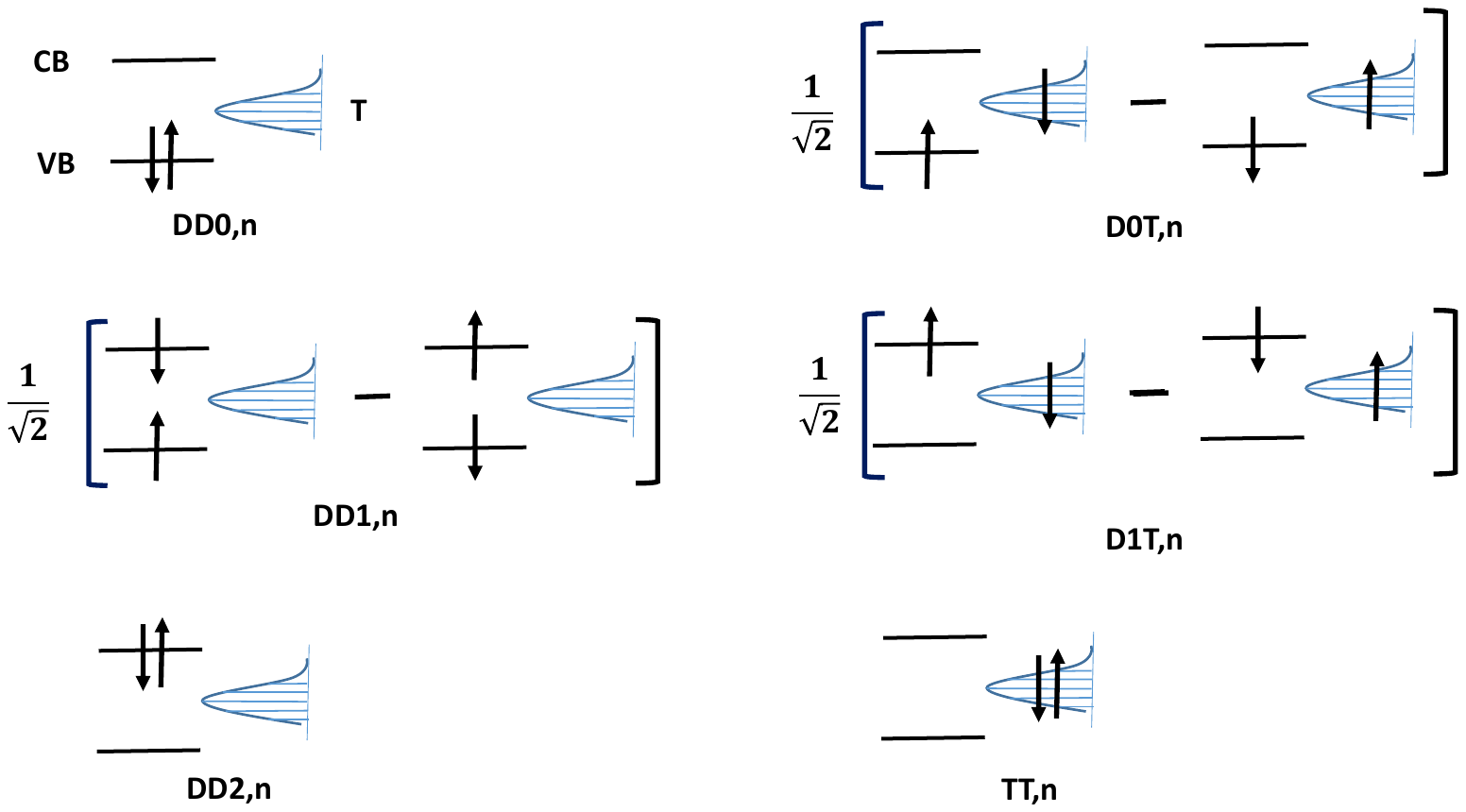}
\caption{ Scheme of all the possible singlet configurations of two electrons in a QD connected to the trap T and $n$ phonons. The left column depicts the three singlets having both electrons in the dot. The right column depicts the two singlets having one electron in the dot and the other in the trap (top and middle) 
and the state with the two electrons in the trap (bottom).
The states are named according to Eq.(7).
} 
\label{esq1}
\end{figure}

\subsection{ Two electrons}

We start considering the simplest system consisting of two electrons. 
Note that the Hamiltonian of Eq. (\ref{Ham}) commutes with  $S^2$ and  $S_z$, $S$ being the total spin and $S_z$ its z-component. Hence $S$ and $S_z$ are conserved, and the singlet and the triplet states of the two-electron system do not mix. 
Before photoexcitation, the system is in its lower energy state that is a singlet. Since light absorption conserves spin, upon photoexcitation the system has to be excited to a singlet state and subsequently evolve in time within the singlet subspace of the Hilbert space.
The states will written in the number representation 

\begin{equation}
|n_{0 \uparrow} n_{0 \downarrow}; n_{1 \uparrow} n_{1 \downarrow}; n_{T \uparrow} n_{T \downarrow}\rangle \otimes |n \rangle,
\label{basis-3}
\end{equation} 
with $n_{0 \sigma}$, $n_{1 \sigma}$ and $n_{T \sigma}$ being the occupation numbers of electrons of spin $\sigma$ in the single-particle levels describing the VB, the CB and the trap, respectively and $n$ being the phononic number. Using this representation, 
the singlet states of the two-electron system that are coupled by $H_{hop}$ and, consequently, evolve in time are

\begin{eqnarray}
|\varphi_{DD0,n}\rangle &= &|1 1; 0 0; 0 0\rangle \otimes|n \rangle,  \nonumber \\
|\varphi_{DD1,n}\rangle &= &\frac{1}{\sqrt 2}(|1 0; 0 1; 0 0\rangle - |0 1; 1 0; 0 0\rangle) \otimes|n \rangle,  \nonumber \\ 
|\varphi_{DD2,n}\rangle &= &|0 0; 1 1; 0 0\rangle \otimes|n \rangle,  \nonumber \\
|\varphi_{D0T,n}\rangle &= &\frac{1}{\sqrt 2}(|1 0; 0 0; 0 1\rangle - |0 1; 0 0 ;1 0\rangle) \otimes|n \rangle,  \nonumber \\
|\varphi_{D1T,n}\rangle &= &\frac{1}{\sqrt 2}(|0 0; 1 0; 0 1\rangle - |0 0; 0 1; 1 0\rangle) \otimes|n \rangle.  \nonumber \\  
|\varphi_{TT,n}\rangle &= &|0 0; 0 0; 1 1\rangle \otimes|n \rangle,  \nonumber \\ 
\label{Singlets}
\end{eqnarray}

These states are schematically drawn in Fig. \ref{esq1}. 
The first three lines of Eq. (\ref{Singlets}) describe the two electrons in the VB (state $DD0,n$), one in the VB and the other in the CB (state $DD1,n$)  
and the two in the CB (state $DD2,n$) (first column in Fig. \ref{esq1}) and $n$ phonons. The fourth and fifth lines represent the two singlet states having one electron in the VB or in the CB and the other electron in the trap and $n$ phonons, states $D0T,n$ and $D1T,n$ respectively, (second column in Fig. \ref{esq1}, top and middle).  
The last line represents the state with the two electrons in the trap and $n$ phonons, (state $TT,n$ in the second column in Fig. \ref{esq1}, bottom). 

The time-dependent wavefunction is then written as a linear combination of these basis states

\begin{eqnarray}
|\psi(t) \rangle = \sum_{n=0}^{\infty}&[&a_{DD0,n}(t)|\varphi_{DD0,n}\rangle +a_{DD1,n}(t)|\varphi_{DD1,n}\rangle +a_{DD2,n}(t)|\varphi_{DD2,n}\rangle \nonumber \\
                                         & &+a_{TT,n}(t)|\varphi_{TT,n}\rangle +a_{D0T,n}(t)|\varphi_{D0T,n}\rangle+a_{D1T,n}(t)|\varphi_{D1T,n}\rangle],
\label{psiSinglets-t}
\end{eqnarray}
and the time-dependent Schr\"odinger equation for Hamiltonian Eq. (\ref{Ham}) is projected onto each of them, leading to the following system of coupled linear differential equations 

\begin{eqnarray}
\frac{d a_{DD0,n}(t)}{dt}&=&-i(2\epsilon_0+n\omega_0)a_{DD0,n}(t)-i \sqrt 2 V_{VB} a_{D0T,n}(t), \nonumber \\
\frac{d a_{DD2,n}(t)}{dt}&=&-i(2\epsilon_1+n\omega_0)a_{DD2,n}(t)-i \sqrt 2 V_{CB} a_{D1T,n}(t), \nonumber \\
\frac{d a_{DD1,n}(t)}{dt}&=&-i(\epsilon_0+\epsilon_1+n\omega_0)a_{DD1,n}(t)-i V_{VB}a_{D0T,n}(t)-i V_{CB} a_{D1T,n}(t)],  \nonumber \\
\frac{d a_{TT,n}(t)}{dt}&=&-i(2 \epsilon_T+U+n\omega_0)a_{TT,n}(t)-i \sqrt 2 V_{VB} a_{D0T,n}(t)-i \sqrt 2 V_{CB} a_{D1T,n}(t)] \nonumber \\
                        & &-i 2 \lambda \sqrt{n}a_{TT,n-1}(t)-i 2 \lambda \sqrt{n+1}a_{TT,n+1}(t), \nonumber \\ 
\frac{d a_{D0T,n}(t)}{dt}&=&-i(\epsilon_0+\epsilon_T+n\omega_0)a_{D0T,n}(t)-i \sqrt 2 V_{VB} a_{DD0,n}(t)-i \sqrt 2 V_{VB} a_{TT,n}(t)-i V_{CB} a_{DD1,n}(t) \nonumber \\
                         & &-i \lambda \sqrt{n}a_{D0T,n-1}(t)-i \lambda \sqrt{n+1}a_{D0T,n+1}(t), \nonumber \\
\frac{d a_{D1T,n}(t)}{dt}&=&-i(\epsilon_1+\epsilon_T+n\omega_0)a_{D1T,n}(t)-i \sqrt 2 V_{CB} a_{DD2,n}(t)-i \sqrt 2 V_{CB} a_{TT,n}(t)-i V_{VB} a_{DD1,n}(t) \nonumber \\
                         & &-i \lambda \sqrt{n}a_{D1T,n-1}(t)-i \lambda \sqrt{n+1}a_{D1T,n+1}(t). \nonumber \\                      
\nonumber \\
\label{equSinglets-t}
\end{eqnarray}

Before photoexcitation the system is in its lower energy state with $n=0$ phonons and at $t=0$ at least one electron is photoexcited to the CB.
Consequently, we have three possible initial conditions.
If the trap is initially empty and the two electrons start in the dot at $t=0$, we have two possible initial states with zero phonons, depicted in the middle and bottom panels in the left column 
 of Fig. \ref{esq1}, the corresponding initial conditions thus being
$a_{DD1,n}(t=0)=\delta_{n,0}$ and $a_{DD2,n}(t=0)=\delta_{n,0}$, respectively.
If the trap is initially occupied with one electron, the other electron being in the CB and $n=0$ phonons, the initial state is the one in the middle panel in the right column of Fig. \ref{esq1} 
and the initial condition is $a_{D1T,n}(t=0)=\delta_{n,0}$.
In each case we solve the set of Eqs. (\ref{equSinglets-t}) with the corresponding initial condition, yielding 
 the occupancies of the VB, CB and trap as
 
\begin{equation}
n_{VB}(t)=\langle \psi(t)| \sum_{\sigma} \hat n_{0,\sigma}|\psi(t)\rangle=  \sum_{n=0}^{\infty} [2| a_{DD0,n}(t)|^2+|a_{D0T,n}(t)|^2+|a_{DD1,n}(t)|^2] ,
\end{equation}

\begin{equation}
n_{CB}(t)=\langle \psi(t)| \sum_{\sigma} \hat n_{1,\sigma}|\psi(t)\rangle=  \sum_{n=0}^{\infty} [2| a_{DD2,n}(t)|^2+|a_{D1T,n}(t)|^2+|a_{DD1,n}(t)|^2] ,
\end{equation}
and

\begin{equation}
n_{T}(t) =\langle \psi(t)| \sum_{\sigma} \hat n_{T,\sigma}|\psi(t)\rangle= \sum_{n=0}^{\infty} [2| a_{TT,n}(t)|^2+|a_{D0T,n}(t)|^2+|a_{D1T,n}(t)|^2] ,
\end{equation}
respectively. The correct normalization of the wavefunction Eq. (\ref{psiSinglets-t})
ensures conservation of the number of electrons at all times, $n_{VB}(t)+n_{CB}(t)+n_{T}(t)=2$, for any of the initial conditions.

\begin{figure}[htbp]
\centering
\includegraphics[width=100mm]{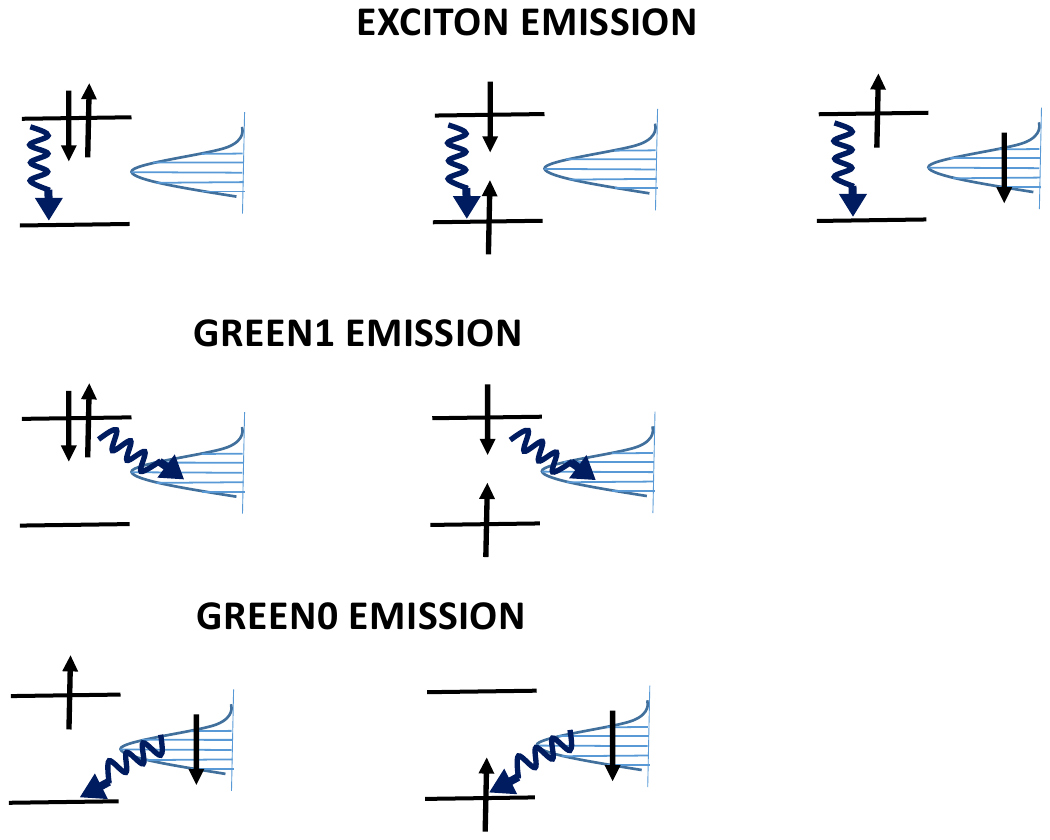}
\caption{Scheme of all the possible states of the two-electron system yielding to light emission of different energies represented by the wavy lines. Only one half of the singlet states $DD1,n$ and $D1T,n$ is drawn for simplicity.
First row: exciton emission, second row: green1 emission and third row: green0 emission.
} 
\label{esq2}
\end{figure}

Light emission of different energies is in principle possible at any time, since the system gets a probability of being in any of its possible states depicted in
Fig. \ref{esq1}. 
There can be exciton emission, coming from deexcitation of one electron from the CB to the VB, with energy $\omega_{exc}=\epsilon_{CB}-\epsilon_{VB}$, or trap luminescence. 
Two lines of trap luminescence, that we call green1 and green0, are possible in this model. The green1 line, with energy $\omega g_1=\epsilon_{CB}-\tilde \epsilon_{T}$
appears if an electron in the CB radiatively decays to the trap level, while the green0 emission, with energy $\omega g_0=\tilde\epsilon_{T}-\epsilon_{VB}$,
appears when an electron in the trap radiatively decays to the VB. The possible states of the two electron system leading to the different lines 
are schematically depicted in Fig. \ref{esq2}.
We do not attempt here to calculate full light emission intensities and line shapes because this will require the calculation of the matrix elements 
for the corresponding transitions which will depend on the material system under consideration. Rather we look
at the statistical factors influencing photoluminescence. These are basically defined as the product of the probabilities that the initial and the
final electronic configurations are statistically accessible states for the light emission process, ie,  
the system has to be in the initial excited state and it cannot be in the final deexcited state (Fig. \ref{esq2}). 
The statistical factors are calculated by means of the following argument.

The interaction Hamiltonian for radiative decay is 

\begin{equation}
\hat H_{rad}=\sum_{i=1,2} \hat h_{rad}(i),
\label{hrad}
\end{equation}
with

\begin{equation}
\hat h_{rad}(i)=\frac{e}{mc} (\hat {\bf p_{i}} \cdot {\bf A}(\bf r_i)+{\bf A}(\bf r_i)\cdot \hat {\bf p_{i}}),
\end{equation}
where $\hat {\bf p_{i}}$ is the momentum operator for particle $i$, $i=1,2$ in our two electron system, ${\bf A}(\bf r_i)$ 
is the vector potential of the electromagnetic field at the electronic coordinate $\bf r_i$, $e$ is the electron charge, 
$m$ is the electron mass and $c$ is the speed of light. In the number representation the Hamiltonian of Eq. (\ref{hrad}) reads

\begin{equation}
\hat H_{rad}=\sum_{i=1,2} \hat h_{rad}(i)=\sum_{r,s} \langle \phi_r|\hat h_{rad}|\phi_s\rangle \hat c_{r}^{\dagger} \hat c_{s},
\end{equation} 
where $r,s$ run over all our single-particle states $0 \sigma$, $1 \sigma$ and $T \sigma$, $\phi_r$ and $\phi_s$ being their
corresponding electronic wavefunctions.

Consider exciton emission from the initial state with the two electrons in the conduction band to the final state with an electron the the CB and the other in the VB
(first row in Fig. \ref{esq2}, left)

\begin{eqnarray}
|initial\rangle= |\varphi_{DD2,n}\rangle &= &|0 0; 1 1; 0 0\rangle \otimes|n \rangle,  \nonumber \\
|final\rangle=|\varphi_{DD1,n}\rangle &= &\frac{1}{\sqrt 2}(|1 0; 0 1; 0 0\rangle - |0 1; 1 0; 0 0\rangle) \otimes|n \rangle,
\end{eqnarray}
with the same number of phonons because $\hat h$ does not act on the phonon coordinates. Then the matrix element is

\begin{equation}
M_{exc}=\langle final| \hat H_{rad}|initial\rangle= \frac{2}{\sqrt2} \langle \phi_0|\hat h_{rad}|\phi_1\rangle.
\label{Mexc}
\end{equation}

Since the probability that the system is in the initial estate is $|a_{DD2,n}(t)|^2$ and the probability that it is not in the final state is 
$1-|a_{DD1,n}(t)|^2$, the transition rate for this process is, to first order in perturbation theory

\begin{equation}
R_{exc1}(t)= |\langle \phi_0|\hat h_{rad}|\phi_1\rangle|^2 2 \sum_{n} |a_{DD2,n}(t)|^2 (1-|a_{DD1,n}(t)|^2).
\end{equation}

Next we consider exciton emission from the initial state with an electron the the CB and the other in the VB to the final state with the two electrons in the VB
(first row in Fig. \ref{esq2}, middle)

\begin{eqnarray}
|initial\rangle=|\varphi_{DD1,n}\rangle &= &\frac{1}{\sqrt 2}(|1 0; 0 1; 0 0\rangle - |0 1; 1 0; 0 0\rangle) \otimes|n \rangle, \nonumber \\
|final\rangle= |\varphi_{DD0,n}\rangle &= &|1 1; 0 0; 0 0\rangle \otimes|n \rangle,  \nonumber \\
\end{eqnarray}
with the same number of phonons. The matrix element is exactly Eq. (\ref{Mexc}) and the corresponding rate is
\begin{equation}
R_{exc2}(t)= |\langle \phi_0|\hat h_{rad}|\phi_1\rangle|^2 2 \sum_{n} |a_{DD1,n}(t)|^2 (1-|a_{DD0,n}(t)|^2).
\end{equation}

Radiative deexcitation from the initial state with one electron in the CB and the other in the trap to the final state with one electron in the VB and the other in the trap
also yields exciton emission (first row in Fig. \ref{esq2}, right). In this case te matrix element is $\frac{M_{exc}}{\sqrt2}$ and the corresponding rate is 

\begin{equation}
R_{exc3}(t)= |\langle \phi_0|\hat h_{rad}|\phi_1\rangle|^2 \sum_{n} |a_{DT1,n}(t)|^2 (1-|a_{DT0,n}(t)|^2).
\end{equation}

The total rate for exciton emission is given by addition of the three independent channels. 
The statistical factor  $P_{exc}(t)$is defined as the quotient of the total rate and 
$|\langle \phi_0|\hat h_{rad}|\phi_1\rangle|^2$

\begin{eqnarray}
P_{exc}(t)=\sum_{n}& & [2 |a_{DD2,n}(t)|^2 (1-|a_{DD1,n}(t)|^2)+2|a_{DD1,n}(t)|^2 (1-|a_{DD0,n}(t)|^2) \nonumber \\
                & & +|a_{DT1,n}(t)|^2 (1-|a_{DT0,n}(t)|^2)]. \nonumber  \\
\label{Pexc}
\end{eqnarray}

When deriving the statistical factors for green1 and green0 emissions we have to take into account 
that the state with two electrons in the trap is not statistically accessible as final state for green1 or initial state for green0 emissions, 
since the corresponding energy transfers will be $\epsilon_{CB}-(2\tilde \epsilon_{T}+U_{eff})$ or $2\tilde \epsilon_{T}+U_{eff}-\epsilon_{VB}$, respectively. 
The accessible states for green1 and green0 emissions are depicted in the second and third rows of Fig. \ref{esq2}, respectively.
Then, proceeding in analogous way, we obtain their statistical factors as

\begin{equation}
P_{g1}(t)=\sum_{n} [2 |a_{DD2,n}(t)|^2 (1-|a_{DT1,n}(t)|^2)+ |a_{DD1,n}(t)|^2 (1-|a_{DT0,n}(t)|^2)] ,
\label{Pg1}
\end{equation}
and

\begin{equation}
P_{g0}(t)=\sum_{n} [2 |a_{DT0,n}(t)|^2 (1-|a_{DD0,n}(t)|^2)+ |a_{DT1,n}(t)|^2 (1-|a_{DD1,n}(t)|^2)],
\label{Pg0}
\end{equation}
respectively.

Eqs. (\ref{Pexc}), (\ref{Pg1}) and (\ref{Pg0}) assume that light emission is an instantaneous process in which the initial and final accessible states occur at the same time $t$.
Although this is an oversimplification, our results show that the three statistical 
factors are not strongly dependent on time, in general. Also, consideration of the light emission channels will affect the coefficients of the different states in the 
wavefunction expansion but we think that in a negligible way. This is because radiative processes, taking times from nanoseconds to microseconds \cite{JPCB_104_1715_vanDijken,  Efros_Nesbitt, Luminescence_Nanoscale, acsnano}, are generally much slower than electronic transitions, taking times of picoseconds \cite{ JPCC_116_20633_cohn, JACS_Schimpf, ChemMat_duFosse, Monreal_PRB}, which means that radiative deexcitation is a much less probable process that electronic relaxations. 
For these reasons we think that the present calculation is relevant.


\subsection{ Three electrons}

The possible configurations of three electrons in our system  with z-component of the total spin $S_z= \frac{1}{2}$,  
in the representation of Eq. (\ref{basis-3}), are

\begin{eqnarray}
|\varphi_{DD0,n}\rangle &= &|1 1; 1 0; 0 0\rangle \otimes|n \rangle,  \nonumber \\
|\varphi_{DD1,n}\rangle &= &|1 0; 1 1; 0 0\rangle \otimes|n \rangle,  \nonumber \\
|\varphi_{D0TT,n}\rangle &= &|1 0; 0 0; 1 1\rangle \otimes|n \rangle,  \nonumber \\
|\varphi_{D1TT,n}\rangle &= &|0 0; 1 0; 1 1\rangle \otimes|n \rangle.  \nonumber \\
|\varphi_{D00T,n}\rangle &= &|1 1; 0 0; 1 0\rangle \otimes|n \rangle,  \nonumber \\
|\varphi_{D11T,n}\rangle &= &|0 0; 1 1; 1 0\rangle \otimes|n \rangle,  \nonumber \\
|\varphi_{a,n}\rangle &= &|1 0; 0 1; 1 0\rangle \otimes|n \rangle,  \nonumber \\
|\varphi_{b,n}\rangle &= &|0 1; 1 0; 1 0\rangle \otimes|n \rangle,  \nonumber \\
|\varphi_{c,n}\rangle &= &|1 0; 1 0; 0 1\rangle \otimes|n \rangle,  \nonumber \\
\label{basis-3e}
\end{eqnarray}

\begin{figure}[htbp]
\centering
\includegraphics[width=100mm]{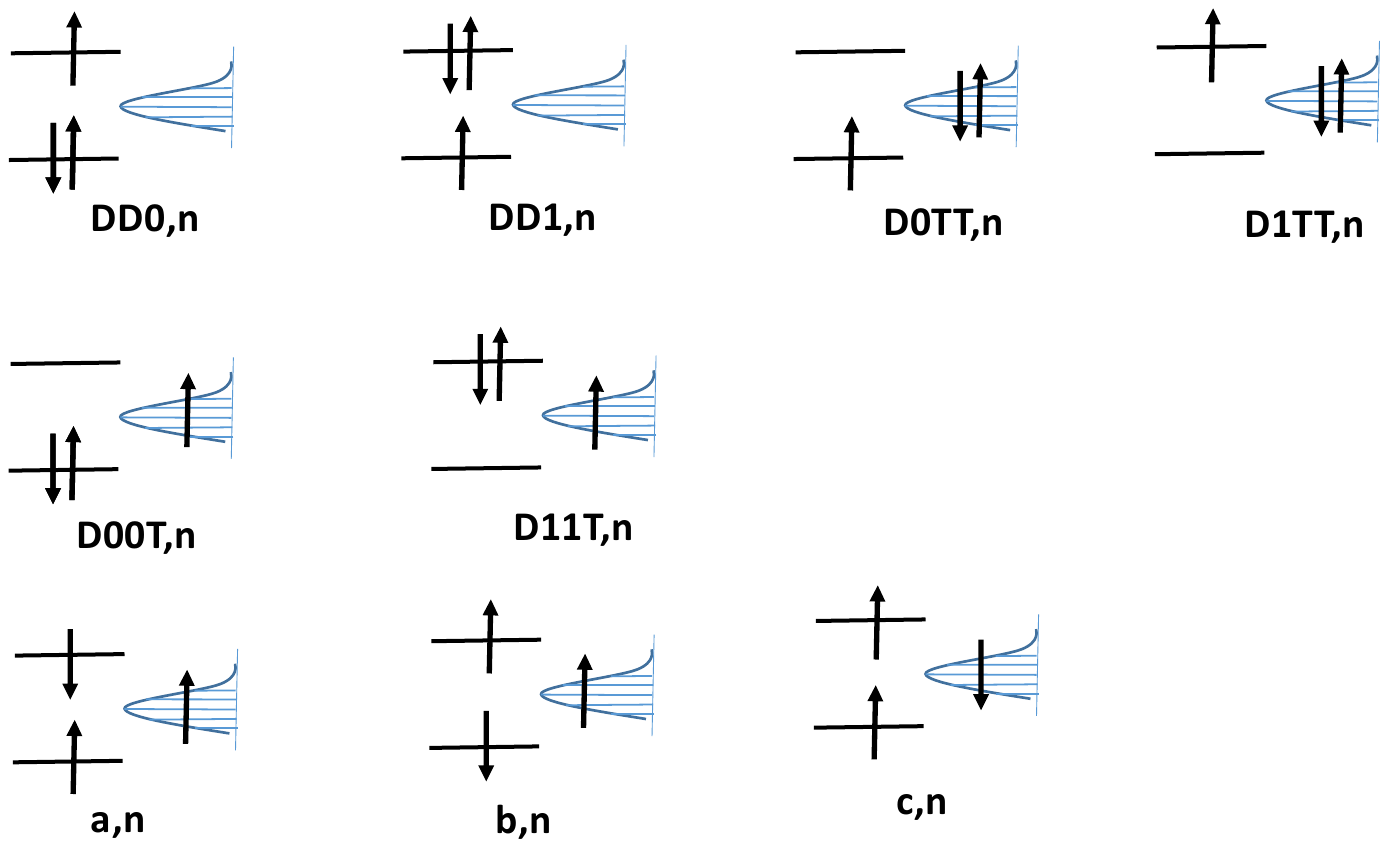}
\caption{Scheme of all the possible configurations of three electrons in a QD connected to the trap T. The states are named according to Eq. (25).
} 
\label{esq3}
\end{figure}

In Eq. (\ref{basis-3e}) the first two lines represent the states of three electrons in the dot, the second and third lines describe the states having two electrons in the trap and one electron in the dot, these four states are drawn in the first row of Fig. \ref{esq3} from left to right, respectively.
 The following five lines describe the states of two electrons in the dot and one in the trap and are depicted in the second and third rows in Fig. \ref{esq3}.
These nine states are coupled by $H_{hop}$ and should be combined in the time dependent wavefunction as 

\begin{eqnarray}
|\psi(t) \rangle = \sum_{n=0}^{\infty}&[&a_{DD0,n}(t)|\varphi_{DD0,n}\rangle +a_{DD1,n}(t)|\varphi_{DD1,n}\rangle \nonumber \\
                                     & &+a_{D00T,n}(t)|\varphi_{D00T,n}\rangle+a_{D11T,n}(t)|\varphi_{D11T,n}\rangle \nonumber \\
                                     & &+a_{a,n}(t)|\varphi_{a,n}\rangle+a_{b,n}(t)|\varphi_{b,n}\rangle+ a_{c,n}(t)|\varphi_{c,n}\rangle \nonumber \\
                                     & &+a_{D0TT,n}(t)|\varphi_{D0TT,n}\rangle + a_{D1TT,n}(t)|\varphi_{D1TT,n}\rangle],
\label{psi3e-t}
\end{eqnarray}

The system of the nine coupled differential equations we have to solve is given in the Appendix A.
If the trap is empty at $t=0$, we have two possible initial states having at least one electron in the CB, namely 
$|\varphi_{DD0,n=0}\rangle$ or $|\varphi_{DD1,n=0}\rangle$. 
If the trap is initially occupied with one electron, the possible initial states having at least one electron in the CB are 
$|\varphi_{D11T,n=0}\rangle$, $|\varphi_{a,n=0}\rangle $, $|\varphi_{b,n=0}\rangle$ or $|\varphi_{c,n=0}\rangle$, see Figure \ref{esq3}.
After solving the system of differential equations with the
corresponding initial condition, we calculate the occupancies of the VB, CB and trap as

\begin{equation}
n_{VB}(t)=\sum_{n=0}^{\infty}[2|a_{DD0,n}(t)|^2+|a_{DD1,n}(t)|^2+2|a_{D00T,n}|^2+|a_{a,n}(t)|^2+|a_{b,n}(t)|^2+|a_{c,n}(t)|^2],
\end{equation}

\begin{equation}
n_{CB}(t)=\sum_{n=0}^{\infty}[|a_{DD0,n}(t)|^2+2|a_{DD1,n}(t)|^2+2|a_{D11T,n}|^2+|a_{a,n}(t)|^2+|a_{b,n}(t)|^2+|a_{c,n}(t)|^2],
\end{equation}
and

\begin{equation}
n_{T}(t)=\sum_{n=0}^{\infty}[2|a_{D0TT,n}(t)|^2+2|a_{D1TT,n}(t)|^2+|a_{D00T,n}|^2+|a_{D11T,n}|^2+|a_{a,n}(t)|^2+|a_{b,n}(t)|^2+|a_{c,n}(t)|^2],
\end{equation}
respectively. Statistical factors for exciton, green1 and green0 emissions are calculated in the same way as in Eqs. (\ref{Pexc}), (\ref{Pg1}) and (\ref{Pg0}), respectively. 

\subsection{Values of the parameters}

The energy gap of wide band gap bulk semiconductors ranges typically between 1.5 eV (GaAs) and 3.5 eV (ZnO). 
The energy gap of nanoparticles increases with decreasing size so the energy
gap of spheres of ca. 2 nm in radius will range between 2 eV (GaAs) and 4 eV (ZnO) \cite{Brus_JCP}. 
Considering the energy level of the trap in the middle of the gap, $\tilde \epsilon_T$ will be between 1 eV and 2 eV.
The choice of the phonon energy is more difficult. The experimental photoluminesce line shapes of a variety of defects in bulk semiconductors 
can be fitted with phonon energies on the order of 25-50 meV \cite{Janotti_PRL}, but it could be different for nanoparticles and it can also depend
on whether the defect is in its volume or at its surface. If the trap were a molecule, vibrational frequencies are in the range of 50 meV to 0.5 eV. 
Using these figures, minimum and maximum values of $\frac{E_g}{\omega_0}$ are 4 and 80 respectively. Here we will use a value in the middle of that range, $\frac{E_g}{\omega_0}= 20, 40$ and
$\frac{\tilde \epsilon_T}{\omega_0}= 10, 20$.
Similar uncertainties arise for choosing the hopping parameters. Typical values of this parameter in band structure calculations are of a few eV.
For $V=1$eV, $V$ will be in the range between $2 \omega_0$ and $20 \omega_0$ for the range of the phonon energies given above. 
Here we will use the smaller value $V_{CB}=\omega_0$ because
one should expect the hopping parameter to be smaller at a low coordination site with a very localized state. 
Nevertheless we have checked that the main conclusions of this work do not change if we increase or decrease the hopping $V_{CB}$ by a factor of 2 
(examples are shown in Figs. \ref{figA3} and  \ref{figA4} in Appendixes D and E, respectively). 
The values of $\frac{\lambda}{\omega_0}$ in the fittings in \cite{Janotti_PRL} are between 3 and 4, 
and values between 1.5 and 3 are also found in other systems \cite{Nelson}. However, values of $\frac{\lambda}{\omega_0} \leq 1$ can be appropriate for shallow traps as in Ref. \cite{acsnano}. In this paper we will use 
several values of $\lambda$, $0 \le \frac{\lambda}{\omega_0} \leq 4$.
The value of $U$ will depend on the atomistic nature of the trap and its surroundings so we perform calculations for several values of 
$U_{eff}$ either negative (effective attraction between two electrons in the trap) or positive (effective repulsion).

In the calculations, we take $\omega_0$ as unit of energy,  $\epsilon_{VB}=0$ and $V_{CB}=\omega_0$. 
The trap level $\tilde \epsilon_{T}$ will be well within the band gap $E_g$, 
$\epsilon_{VB} \ll \tilde \epsilon_T \ll \epsilon_{CB}$, and we will explore the dependence of the occupancies and statistical factors with 
$\tilde \epsilon_T$, $E_g$, $V_{VB}$ and $U_{eff}$. 
The time-dependent equations are solved using a standard Runge-Kutta method. The value of the time step depends on the parameters and the calculations are checked by ensuring the correct normalization of all the wavefunctions with an accuracy better than 1$\%$, $ \forall t$. The number of phonons that should be included in the calculation obviously depends on $\lambda$ and we need to include typically 100 phonons for $\lambda=3,4$. 
This indicates that a large number of phonons are virtually involved in the process.
The calculations run up to a maximum time $\omega_0 t_{max}=200$ which is usually enough to have converged results. 

\section{Results and Discussion}

\begin{figure}[htbp]
\centering
\includegraphics[width=80mm]{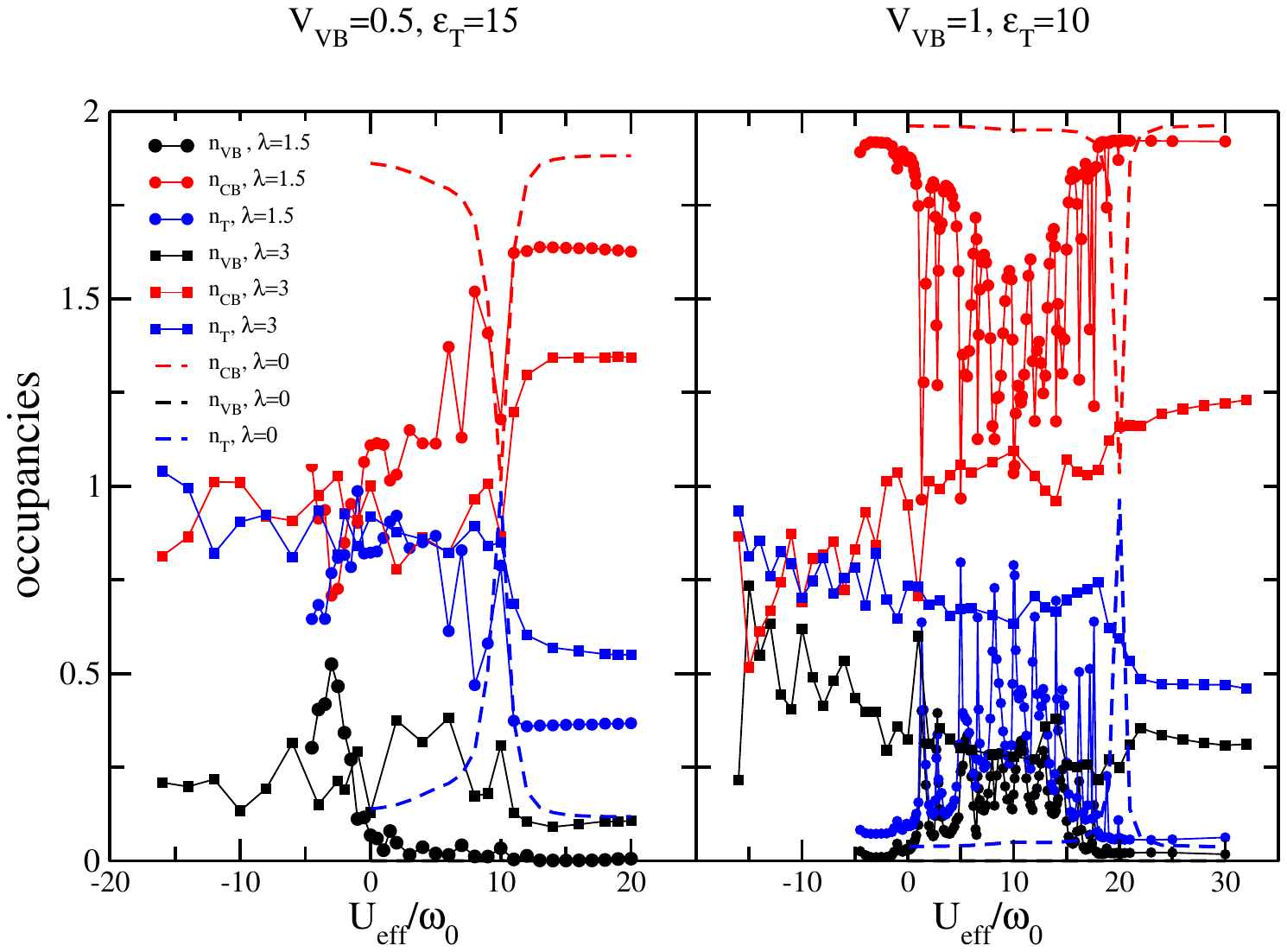}
\caption{\label{fig1} Time averaged occupancies for a system of two electrons in a QD with $E_{g}=20 \omega_0$, as a function of $U_{eff}$ and three values of $\lambda$. 
The initial state has two electrons in the CB ($DD2,n=0$ in Fig. \ref{esq1}). 
Right panel is for a perfectly symmetric case with $\tilde \epsilon_{T}=10 \omega_0$, $V_{CB}= V_{VB}=\omega_0$.  
Left panel is for an asymmetric case with $\tilde \epsilon_{T}=15 \omega_0$, $V_{CB}= \omega_0$ and $V_{VB}=0.5 \omega_0$. 
Black symbols: $n_{VB}$, red symbols: $n_{CB}$ and blue symbols: $n_{T}$. Dashed lines without symbols: $\lambda=0$, dots: $\lambda=1.5\omega_0$ and
squares: $\lambda=3 \omega_0$
.} 
\end{figure}

Although this paper focuses on the dynamical properties of the electron-phonon system, it is illustrative to start discussing
the time-averaged values of the electron occupancies defined as
 
\begin{equation}
<n_{i}>=\frac{1}{t_{max}} \int_{0}^{t_{max}} dt n_{i}(t),
\end{equation}
for $i=$ VB, CB and T.
 
Fig. \ref{fig1} shows the time averaged occupancies of the CB, VB and trap as a function of $U_{eff}$ for $\lambda=0$, $\lambda=1.5\omega_0$ and
 $\lambda=3 \omega_0$. 
The QD has $E_{g}=20 \omega_0$ and the initial state has two electrons in the CB ($DD2,n=0$ in Fig. \ref{esq1}). 
The right panel is for a perfectly symmetric case in which the trap level is exactly in the middle of the gap, $\tilde \epsilon_{T}= 10 \omega_0$ and  
$V_{CB}= V_{VB}=\omega_0$.
The left panel is for the asymmetric case with $\tilde \epsilon_{T}=15 \omega_0$, $V_{CB}= \omega_0$ and $V_{VB}=0.5 \omega_0$. In this case we 
have chosen $V_{VB}$ smaller than $V_{CB}$ because the trap level is farther away from the VB than from
the CB and we expect the overlap to be smaller. 
In the absence of electron-phonon interaction ($\lambda=0$), $n_{VB}$ is negligible for all values of $U_{eff}$ and $n_T$ is appreciable only near the value of $U_{eff}$ for which the energy of the two electrons in the CD, $2\epsilon_{CB}$,
equals the energy of the two electrons in the trap, $2 \tilde \epsilon_{T}+U_{eff}$. Then both electrons jump together back and forth between CB and trap periodically in both symmetric and asymmetric cases. 
Consideration of electron-phonon interaction introduces trap sub-levels near the VB and the CB that possibilitates efficient electron transfer between levels,
the amount of which depends on the system parameters. 
For $\lambda=1.5 \omega_0$ (dots), $n_{VB}$ is appreciable only for $U_{eff} \lesssim 0$ in the asymmetric case while it shows 
very narrow peaks superimposed on a continuum with a
maximum of 20 $\%$  at $U_{eff} \simeq 10\omega_0$ in the symmetric case. 
This is because the weight of the trap sub-levels near the VB is smaller in the asymmetric case than in the symmetric one, 
the trap level being farther away from the VB in the former case, and the hopping parameter to the VB being also smaller. 
On the contrary, $n_T$ is much larger for the asymmetric case than for the symmetric one because the weight of the trap sub-levels near the CB is larger in the former case, the hopping parameter to the CB having the same value in both.
Increasing $\lambda$ to $\lambda=3 \omega_0$ (squares) increases the values of $n_{VB}$ in both cases while $n_T$ does not change much in the asymmetric case but still increases in the symmetric one. 
Note in the symmetric case that the initially empty VB (two holes) can get ca. 0.5 electrons for many values of $U_{eff}$ due to strong electron-phonon interaction. 
Also note how for $U_{eff} \lesssim 0$  $n_T$ and $n_{CB}$ get nearly equal values with $n_{VB}$ taking also
its larger values, in both symmetric and asymmetric cases. This is because the effective attraction of the two electrons in the trap favors its double occupancy and then electron transfer to the VB. 
We will show below that it also affects the electron-phonon dynamics, showing signatures of strong electron correlations that are masked in the time-averaged values of the occupancies.  
We will present several examples having similar values of the time-averaged occupancies but showing very different dynamics.

\begin{figure}[htbp]
\centering
\includegraphics[width=80mm]{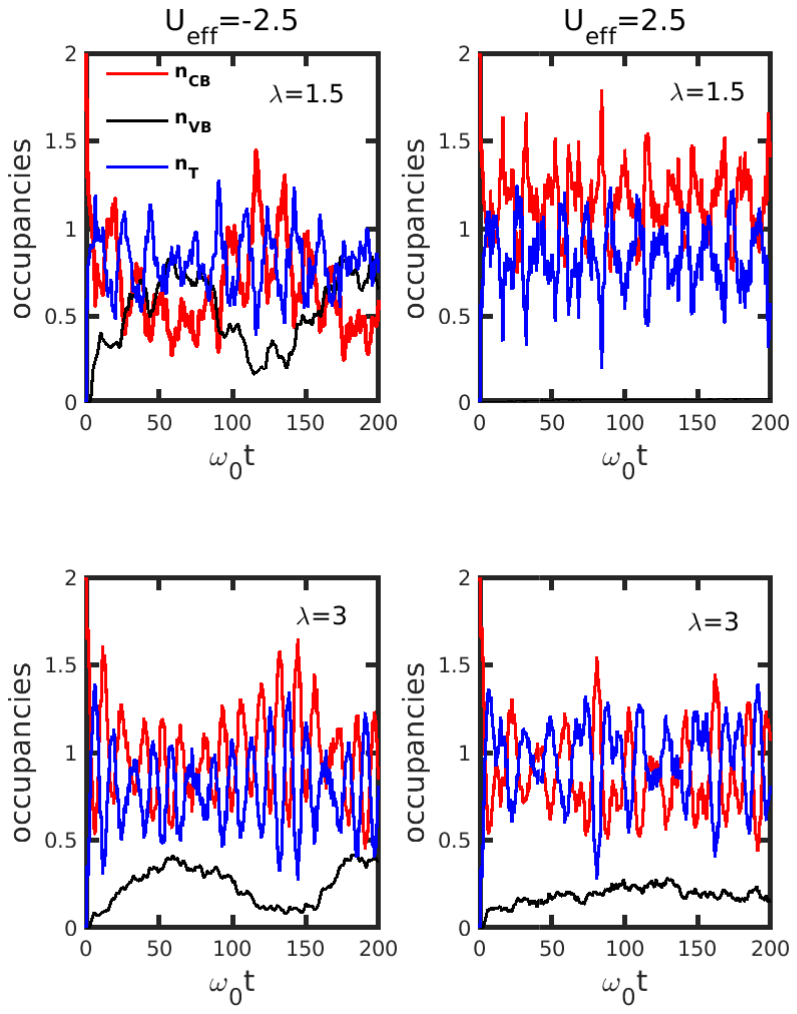}
\caption{\label{fig2} Time dependent occupancies $n_{CB}(t)$ (red lines), $n_{VB}(t)$ (black lines) and $n_{T}(t)$ (blue lines)
for the asymmetric case, $E_{g}= 20 \omega_0$, $\tilde \epsilon_{T}=15 \omega_0$, $V_{CB}= \omega_0$ and $V_{VB}=0.5 \omega_0$, two values of $\lambda$ and two values of $U_{eff}$. 
The initial state has two electrons in the CB ($DD2,n=0$ in Fig. \ref{esq1}). 
The two upper panels are for $\lambda=1.5 \omega_0$ and two lower panels are for $\lambda=3 \omega_0$.
The two left panels are for $U_{eff}= -2.5 \omega_0$ and the two right panels are for $U_{eff}= 2.5 \omega_0$.
The sign of $U_{eff}$ strongly affects the electron dynamics
.} 

\end{figure}

Figure \ref{fig2} shows the system dynamics for the asymmetric case, $\tilde \epsilon_{T}=15 \omega_0$, $V_{CB}= \omega_0$ and $V_{VB}=0.5 \omega_0$, 
two values of $\lambda$ and two values of $U_{eff}$. The initial state has two electrons in the CB ($DD2,n=0$ in Fig. \ref{esq1}). 
We first note the fast oscillations of $n_{CB}(t)$ and $n_{T}(t)$ of large amplitude and opposite phases appearing in all cases that correspond to the
fast exchange of electrons between the CB and the trap. In contrast, $n_{VB}(t)$ shows a much slower dynamics, either oscillating with a larger period
(left panels) or staying almost constant (right panels). We attribute the slow oscillation to the existence of a trap sublevel in quasi-resonance with the VB level 
that allows an efficient electron transfer from the trap to the VB. We found this kind of behavior when studying the simplest case of a single electron hopping
between a single level QD and the trap, where an approximate analytical solution of the time-dependent Sch\"ordinger equation can be found in some limits. 
In particular, in the strong coupling regime  $\frac{\lambda}{\omega_0} \gg 1$ we showed that the electron hopping is periodic, with a renormalized period via a renormalized hopping parameter,
both depending on $\lambda$.
This explains the differences in intensity and period of the oscillations in $n_{VB}$ shown in the two left panels of Fig. \ref{fig2}, differing in $\lambda$. 
However, this kind of behavior only happens for negative (or positive but small) values of $U_{eff}$, since, 
as stated above, the effective attraction among the electrons favors electron transfer from trap to the VB. 
Actually, we have not found it for large and positive values of $U_{eff}$ in any of the cases investigated in this work.
Also notice on the left upper panel the time regions where the maxima in $n_{VB}(t)$ are the minima in $n_{CB}(t)$ with $n_{T}(t)$ staying almost constant and
$n_{VB} \simeq n_T$.
This is surprising since the value of the hopping parameter to the CB is twice the value of the hopping to the VB.
We think this is a signature of strong electronic correlations that occur for $U_{eff} \leq 0$.
Finally, we point out that these signatures do not always show up in time-averaged values of the levels occupancies, as can be seen when comparing
the two lower panels in Fig. \ref{fig2}, having very different dynamics but nearly equal values of the time-averaged occupancies.

\begin{figure}[htbp]
\centering
\includegraphics[width=80mm]{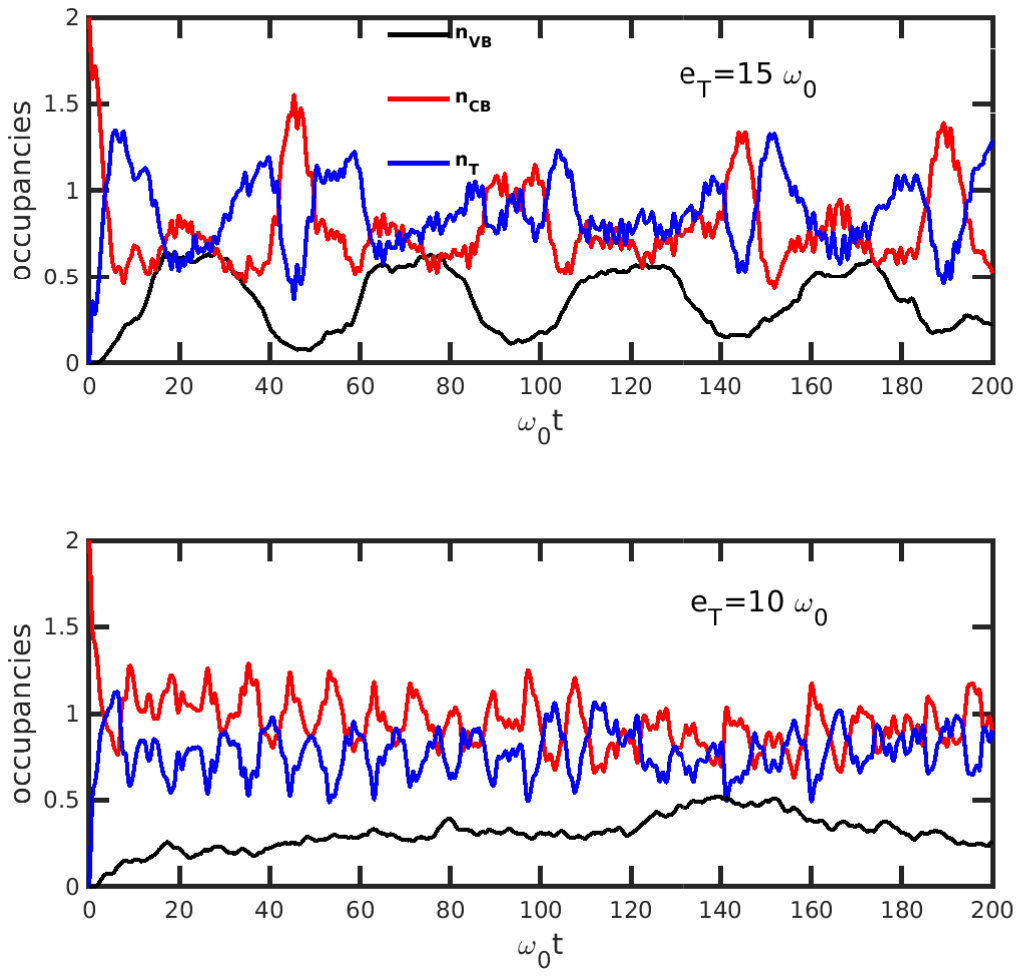}
\caption{\label{fig3} Time dependent occupancies $n_{CB}(t)$ (red lines), $n_{VB}(t)$ (black lines) and $n_{T}(t)$ (blue lines)
for two cases differing only in $\tilde \epsilon_{T}$. 
In both $E_{g}= 20 \omega_0$, $\lambda=3 \omega_0$, $V_{CB}= \omega_0$, $V_{VB}=0.5 \omega_0$ and $U_{eff}= 2 \omega_0$.  
The initial state has two electrons in the CB ($DD2,n=0$ in Fig. \ref{esq1}). 
The upper panel is for $\tilde \epsilon_{T}=15 \omega_0$ and the lower panel is for $\tilde \epsilon_{T}=10 \omega_0$.
In spite of the very different dynamics, the time-averaged values of the three occupancies are almost identical
.} 
\end{figure}

Another example of this behavior is presented in the two panels in Fig. \ref{fig3}, differing only in $\tilde \epsilon_T$, the values of the rest of the parameters being the same. 
Again, the system dynamics looks very different but the time-averaged values of the electron occupancies are almost identical.
As in Fig. \ref{fig2} we find the fast oscillations of $n_{CB}(t)$ and $n_{T}(t)$ with the slow variations of $n_{VB}(t)$ in both cases but the electron dynamics is 
correlated for $\tilde \epsilon_T=15 \omega_0$ (upper panel) and not so much for $\tilde \epsilon_T=10 \omega_0$ (lower panel).
In the upper panel we find the flat regions of time where the three occupancies are constant with nearly equal values, 
characteristic of correlated electronic motion, that in this case happens even for this positive but small value of  $U_{eff}$. As stated above, 
we attribute the slow oscillation in $n_{VB}(t)$ to the existence of a trap sublevel in quasi-resonance with the VB level.  
Since the position of the trap sub-levels depend on $U_{eff}$, a small change in $U_{eff}$ can change the electron dynamics substantially. This is what we see when we only change 
the value of $U_{eff}$ from
$U_{eff}= 2 \omega_0$ in the upper panel of Fig. \ref{fig3} to $U_{eff}= 2.5 \omega_0$ in the right bottom panel of Fig. \ref{fig2}. 
It is interesting to note that in the first $t \lesssim \frac{5}{\omega_0}$s ca 1.5 of the two electrons 
initially in the CB are transferred to the trap, this being an extremely fast process. Later on, a quasi-stationary dynamics is established in which up to 0.5 of the electrons are transferred to the VB in
a slower process. 
The fast transient is actually determined by the hopping parameter $V_{CB}$, that in this calculation equals $\omega_0$,
 as $t \lesssim \frac{5}{V_{CB}}$.  Consequently,
 $t \lesssim 4 $fs for $V_{CB}=1$eV, as estimated in Section II C, but it can be larger for trap states at nanoparticle surfaces with smaller
 hoppings.
 The time for electronic transfer to the VB depends on $\lambda$, as we saw in 
 Fig. \ref{fig2}, and also on $V_{VB}$ as we will see next.  

\begin{figure}[htbp]
\centering
\includegraphics[width=80mm]{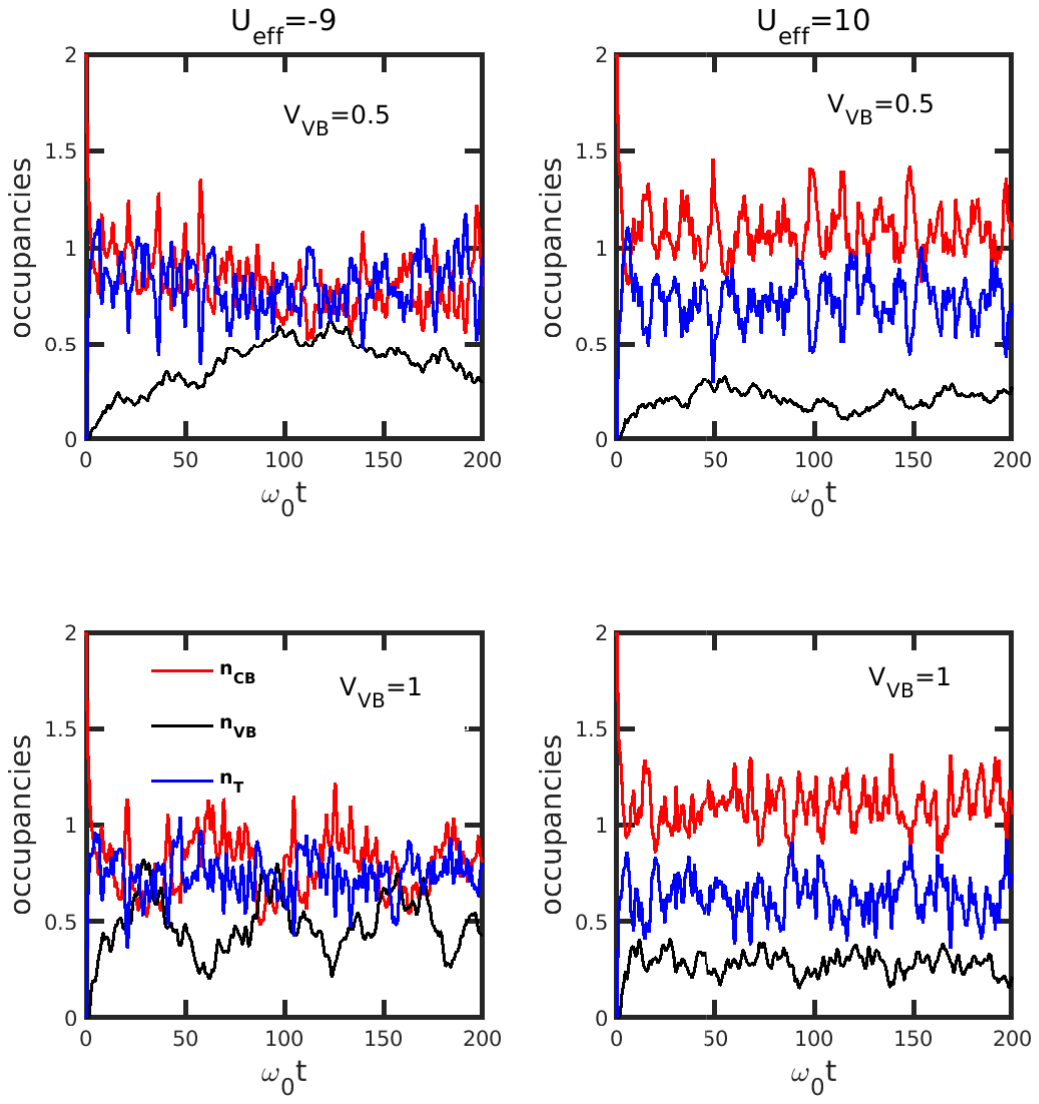}
\caption{\label{fig4} Time dependent occupancies $n_{CB}(t)$ (red lines), $n_{VB}(t)$ (black lines) and $n_{T}(t)$ (blue lines)
for four cases with $\tilde \epsilon_{T}=10 \omega_0$, $E_{g}= 20 \omega_0$, $V_{CB}=\omega_0$ and $\lambda=3 \omega_0$,
differing in $V_{VB}$ and in $U_{eff}$. 
The initial state has two electrons in the CB ($DD2,n=0$ in Fig. \ref{esq1}). 
The two top panels are for $V_{VB}=0.5 \omega_0$ and two bottom panels are for $V_{VB}= \omega_0$.
The two left panels are for $U_{eff}= -9 \omega_0$ and the two right panels are for $U_{eff}= 10 \omega_0$
.} 
\end{figure}

In Figure \ref{fig4} we compare cases with large negative and positive values of $U_{eff}$ and two values of $V_{VB}$.
For $U_{eff}= -9 \omega_0$ (left panels) the correlated electronic motion is evident for both values of $V_{VB}$, 
with $n_{CB}(t) \le 1$ $\forall t$, although the dynamics of $n_{VB}(t)$ 
is slower for the smaller value of $V_{VB}$. (Actually this is one of the slowest we have found along this investigation). 
In particular, notice in the left bottom panel that $n_{T}(t)$ basically presents fast oscillations of small amplitude around its time-averaged value while
the minima of $n_{VB}(t)$ are the maxima of $n_{CB}(t)$ and all the three occupancies have the same value at the maxima of $n_{VB}(t)$.
None of these features is seen at the large and positive $U_{eff}= 10 \omega_0$ (right panels) for any value of $V_{VB}$. 
In these two cases, after a transient time of $t \lesssim \frac{10}{V_{CB}}$, ($V_{CB}=\omega_0$ in these calculations), the three occupancies show small oscillations around their time-averaged values, which are 1.1 electrons in the CB, 0.6 electrons in the trap and
0.3 electrons in the VB,
basically depicting one of the initial electrons staying in the CB, and the other one shared between the trap and the VB,
this quasi-stationary situation being rather independent of $V_{VB}$.
However, for $|U_{eff}| \simeq V_{VB}$
the correlated/uncorrelated dynamics depends on $V_{VB}$. This is illustrated in Figure \ref{figA1} of Appendix B
where we plot the time dependent occupancies for $U_{eff}= \pm \omega_0$ and the same values of $V_{VB}$ as in this Figure.

\begin{figure}[htbp]
\centering
\includegraphics[width=80mm]{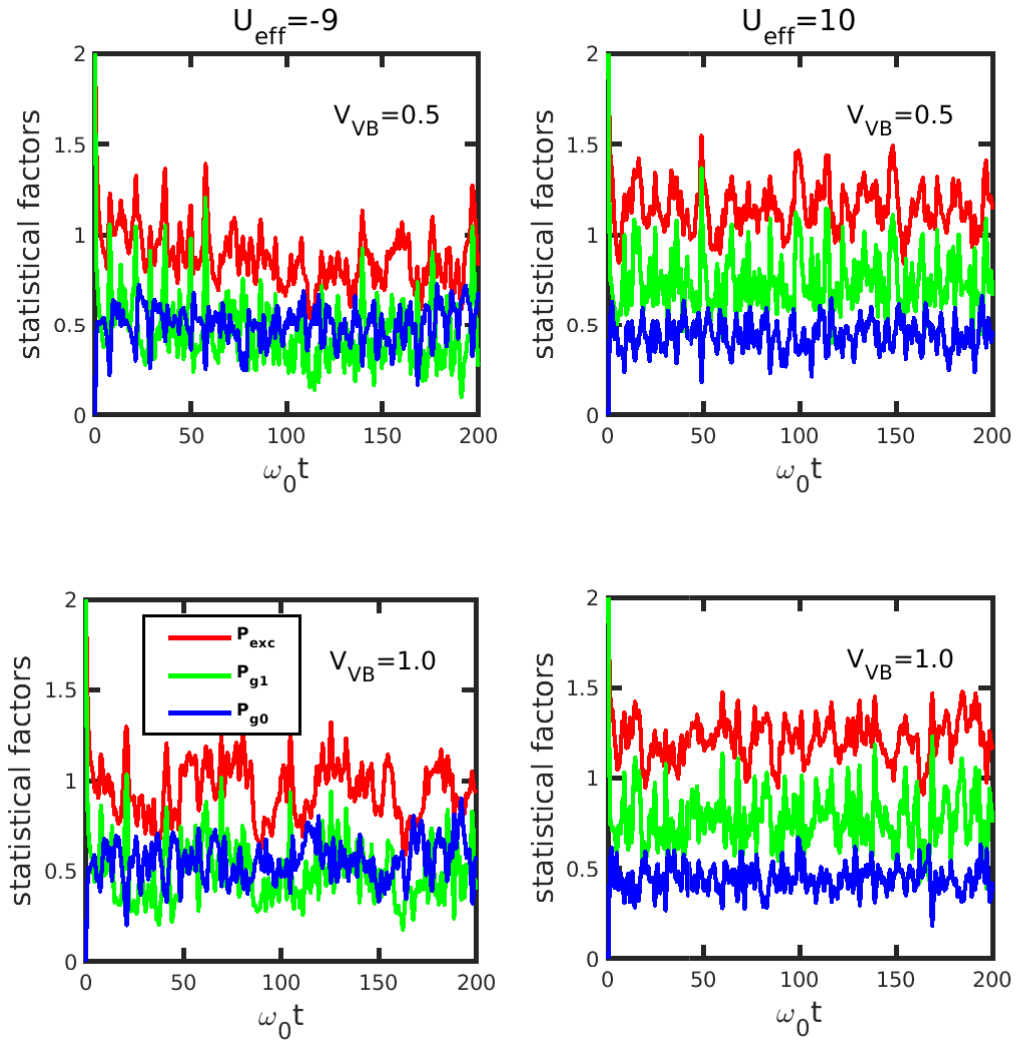}
\caption{\label{fig5} Time dependent light emission statistical factors $P_{exc}(t)$ (red lines), $P_{g1}(t)$ (green lines) and $P_{g0}(t)$ (blue lines) 
for the same four cases as in Fig.\ref{fig4}, all with $\tilde \epsilon_{T}=10 \omega_0$, $E_{g}= 20 \omega_0$ $V_{CB}=\omega_0$ and $\lambda=3 \omega_0$.  
The initial state has two electrons in the CB ($DD2,n=0$ in Fig. \ref{esq1}). 
The two top panels are for $V_{VB}=0.5 \omega_0$ and two bottom panels are for $V_{VB}= \omega_0$.
The two left panels are for $U_{eff}= -9 \omega_0$ and the two right panels are for $U_{eff}= 10 \omega_0$.
In spite of the very different electron dynamics shown in Fig. \ref{fig4},
the light emission factors follow similar patterns
.} 
\end{figure}

In spite of the very different dynamics shown by the electron occupancies, the statistical factors for light emission almost always show identical pattern:
they present the fast oscillations associated to the electronic hopping between the CB and the trap and are quite independent of the motion of the VB electrons.
This is illustrated in Figure \ref{fig5}
displaying the statistical factors for excitonic emission (red lines), green1 emission (green lines) and green0 emission (blue lines) for the same four cases shown in
Fig. \ref{fig4}. $P_{exc}(t)$ is the largest and follows closely $n_{CB}(t)$. $P_{g1}(t)$ also tends to follow $n_{CB}(t)$ but
 $P_{g0}(t)$ cannot be clearly related to any of the occupancies, although it also presents the same fast oscillations. 
The insensitiveness of the statistical factors to the sign of $U_{eff}$ points to a minor effect of electronic correlations
on the light emission properties, 
a possible reason for this is as follows. For negative values of $U_{eff}$, the double-occupied trap state has an important weight in the time-dependent wavefunction,
but this state does not contribute to the green emissions although it strongly affects the electron occupancies, as discussed above, while for positive values of $U_{eff}$ the electronic motion is basically uncorrelated.

\begin{figure}[htbp]
\centering
\includegraphics[width=80mm]{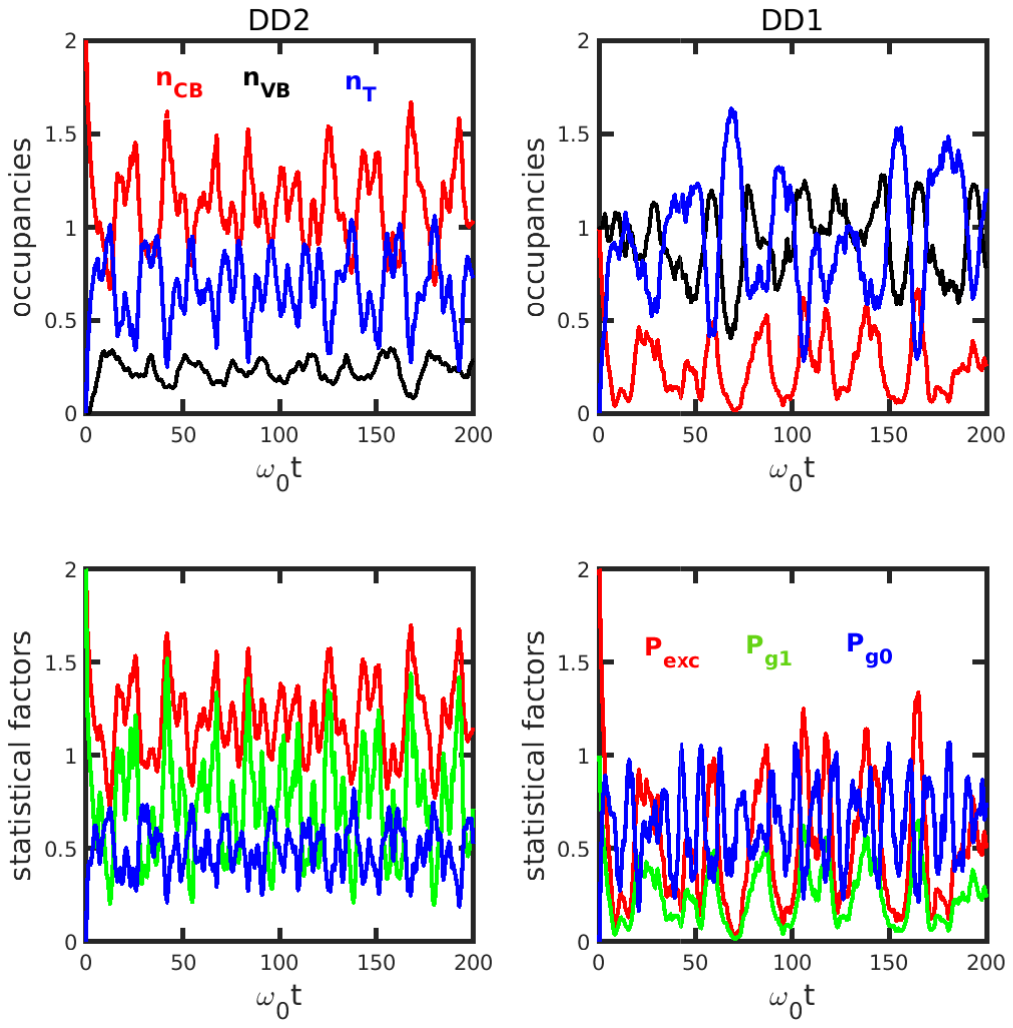}
\caption{\label{fig6} Time dependent occupancies (top panels) and light emission statistical factors (bottom panels) for
$E_{g}= 40 \omega_0$, $\tilde \epsilon_{T}=20 \omega_0$, $V_{VB}= V_{CB}=\omega_0$, $\lambda=4 \omega_0$, $U_{eff}= -2 \omega_0$ and two different initial conditions.
The left panels are for $DD2,n=0$ as initial state and in the right panels the initial state is $DD1,n=0$. 
In the top panels red lines plot $n_{CB}$, blue lines plot $n_{T}$ and black lines plot $n_{VB}$.
In the bottom panels red lines plot $P_{exc}$, green lines plot $P_{g1}$ and blue lines plot $P_{g0}$
.} 
\end{figure}

As we said above, the system dynamics is also dependent on the initial conditions. Figure \ref{fig6} compares occupancies and light emission statistical factors 
for two different initial conditions. The left panels are for $DD2,n=0$ as initial state (two electrons in the CB) while $DD1,n=0$ (one electron in the VB and one electron in the CB) 
is the initial sate in the right panels. 
This calculation is for a symmetric system with a band gap twice the one used in the previous figures,
$E_{g}= 40 \omega_0$, $\tilde \epsilon_{T}=20 \omega_0$, $V_{VB}= V_{CB}=\omega_0$, $\lambda=4 \omega_0$ and $U_{eff}= -2 \omega_0$. We have chosen these values because 
they produce values of time-averaged occupancies similar to the cases shown in previous figures for the symmetric system.
With $DD2,n=0$ as initial state, the occupancies and the light emission factors look very similar to the ones shown in Figs. \ref{fig4} and \ref{fig5}, respectively,
for the same initial state and positive $U_{eff}$, that we discussed above. 
However, with $DD1,n=0$ as initial state we find that the three occupancies present the same oscillatory pattern, 
with $n_{CB}(t)$ and $n_{VB}(t)$ being in phase and $n_{T}(t)$ being out of phase, and the electronic motion is correlated: 
at times when $n_{CB}(t)=0$, $n_{T}(t) \simeq 1.5$ and $n_{VB}(t) \simeq 0.5 $, depicting the simultaneous transfer to the trap of the electron initially in the CB 
and half of the electron initially in the VB. At these times $P_{exc}(t)= P_{g1}(t)=0$, as it should, and only emission of the green0 light is possible. However, the three lines can be emitted at other times with similar intensities. Note in both calculations that the long time 
system dynamics is established after $t \lesssim \frac{10}{V_{CB}}$, as in the previous cases, all having $V_{CB}=\omega_0$.

\begin{figure}[htbp]
\centering
\includegraphics[width=80mm]{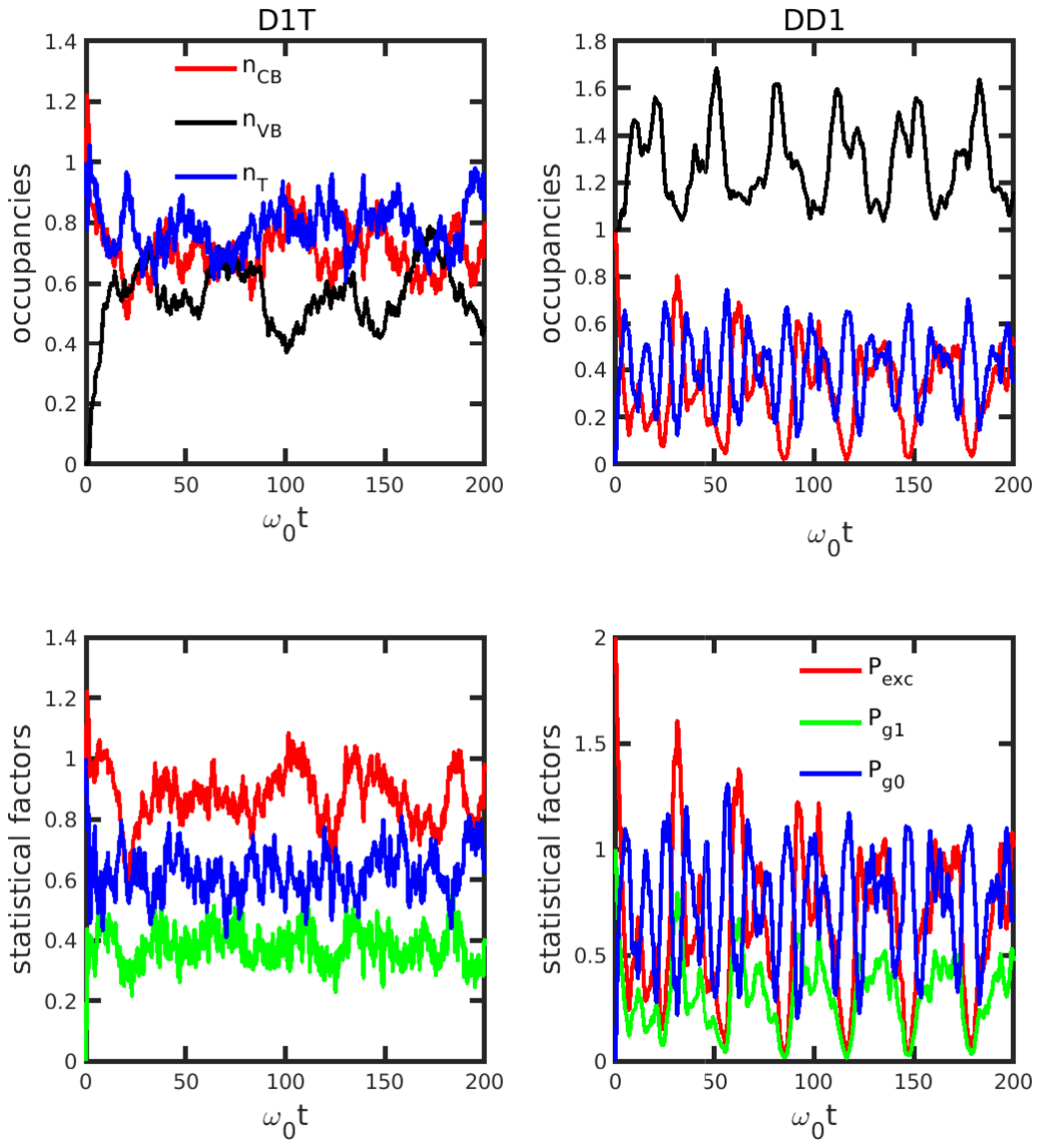}
\caption{\label{fig7} Time dependent occupancies (top panels) and light emission statistical factors (bottom panels) for
$E_{g}= 40 \omega_0$, $\tilde \epsilon_{T}=20 \omega_0$, $V_{VB}= V_{CB}=\omega_0$, $\lambda=4 \omega_0$, $U_{eff}= 2 \omega_0$ 
and two initial conditions differing in the initial occupancy of the trap.
In the left panels the initial state is $D1T,n=0$ (Fig. \ref{esq1}). 
In the right panels the initial state is $DD1,n=0$ (Fig. \ref{esq1}).
In the top panels red lines plot $n_{CB}$, blue lines plot $n_{T}$ and black lines plot $n_{VB}$.
In the bottom panels red lines plot $P_{exc}$, green lines plot $P_{g1}$ and blue lines plot $P_{g0}$
.} 
\end{figure}

Figure \ref{fig7} shows time-dependent occupancies and light emission statistical factors for two initial conditions differing in the occupancy of the trap,
with the same values of the parameters as in Fig. \ref{fig6} except for $U_{eff}= 2 \omega_0$ in this figure. 
In the left panels, with $D1T,n=0$ as initial state, the initially empty VB is quickly filled by CB and trap electrons, in times $t \simeq \frac{10}{V_{CB}}$, 
 and the number of electrons there remains oscillating between 0.4 and 0.7, 
which is a large number in comparison with the other presented cases. 
The three statistical factors, however, present the same fast oscillations as in previous cases. 
In the right panels, where $DD1,n=0$ is the initial state, the electron initially in the VB stays there 
($n_{VB}(t) \ge 1$ $\forall t$) and the electron in the CB jumps back and forth from the CB to the trap and then to the VB, the VB getting 0.6 extra electrons at most. This is an uncorrelated motion (basically only one electron moves) differing from the the correlated one shown in Fig. \ref{fig6} for $U_{eff}=-2 \omega_0$,
but the three statistical factors look very similar in both cases.
We again find that signs of electronic correlations show up in the time-dependent occupancies but not in the time-dependent light emission factors. 
However, we must point out that the magnitude of the light emission factors is influenced by these correlations via the weight of the states in the system wavefunction 
Eq. (\ref{psiSinglets-t}):
a large weight of the double-occupied trap state has to be compensated by a smaller weight of the other states. We find that the excitonic statistical factor $P_{exc}(t)$ 
is systematically the largest while $P_{g1}(t) \geq P_{g0}(t)$ for correlated motion and $P_{g1}(t) \lesssim P_{g0}(t)$ for uncorrelated motion.

\begin{figure}[htbp]
\centering
\includegraphics[width=80mm]{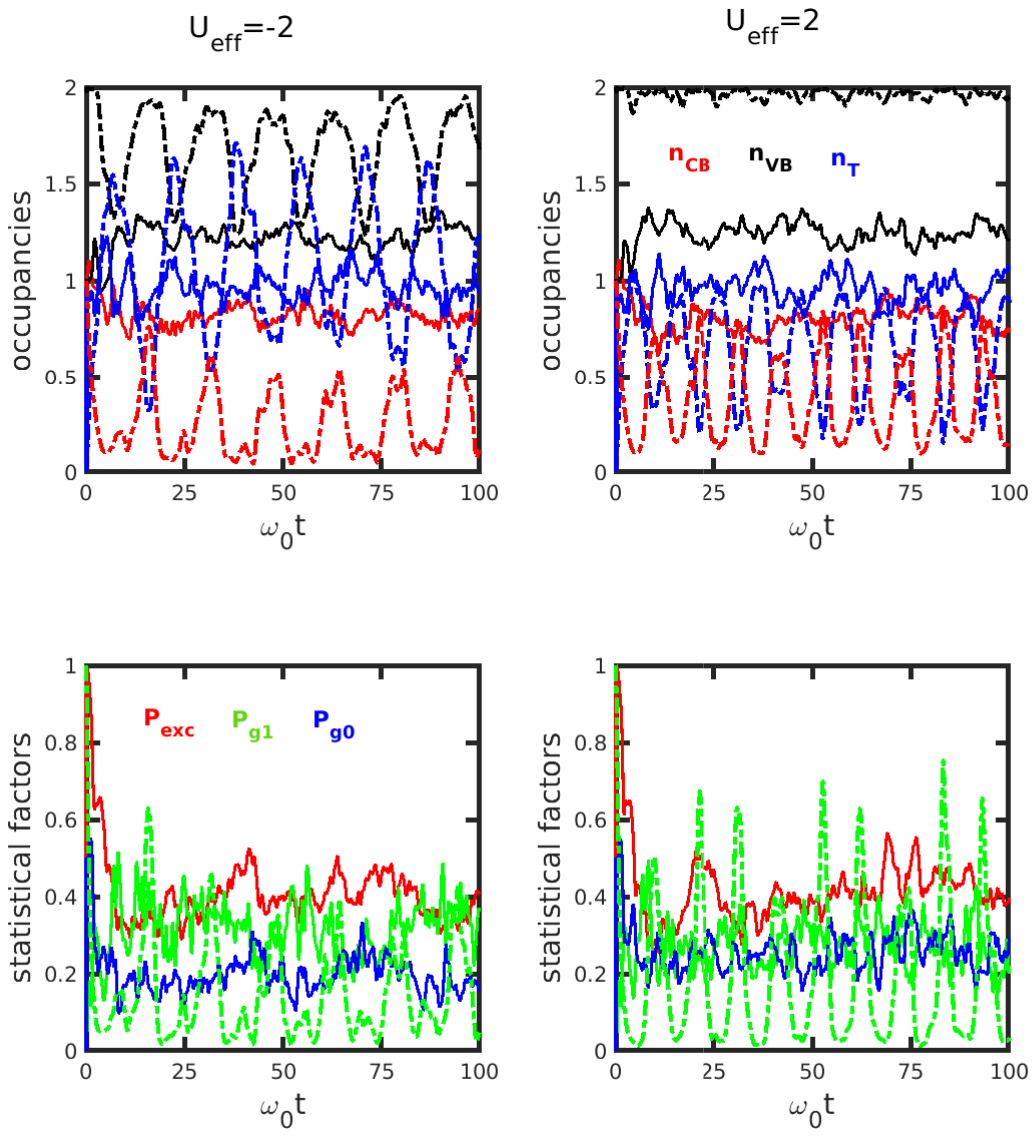}
\caption{\label{fig8} Time dependent occupancies (top panels) and light emission statistical factors (bottom panels) for a three electron system with
$\tilde \epsilon_{T}=10 \omega_0$, $E_{g}= 20 \omega_0$, $V_{VB}= V_{CB}=\omega_0$, $\lambda=3 \omega_0$, two values of $U_{eff}$ 
and two initial conditions differing in the initial occupancy of the trap drawn as: 
continuous lines are for an electron in the trap, another in the CB and the other in the VB (state $a,n=0$ in Fig. \ref{esq3}); 
dashed lines are for initial state with two electrons in the VB and the other in the CB (state $DD0,n=0$ in Fig. \ref{esq3}).
Left panels are for $U_{eff}=-2 \omega_0$ and right panels are for $U_{eff}=2 \omega_0$.
In the top panels red lines plot $n_{CB}$, blue lines plot $n_{T}$ and black lines plot $n_{VB}$.
In the bottom panels red lines plot $P_{exc}$, green lines plot $P_{g1}$ and blue lines plot $P_{g0}$
.} 
\end{figure}

Finally, we present results for three electrons. In Figure \ref{fig8} we compare calculations performed with two initial conditions, one with the trap initially occupied
(continuous lines) and the other with the trap initially empty (dashed lines), 
and two values of $U_{eff}$, one negative (left panels) and the other positive (right panels).  
The case of an initially occupied trap shows a quasi-stationary dynamics, with all the three occupancies and the three statistical light emission factors 
reaching very fast nearly constant values, which are rather independent of $V_{CB}$, in times $t \simeq \frac{10}{V_{CB}}$, this kind of dynamics being insensitive to the sign of $U_{eff}$.
For the three electron system, we have found the same behavior not only for an initially occupied trap but for many other initial conditions as well, an example is shown in Fig. \ref{figA2} in Appendix C. 
In contrast, the case in which the VB is initially double-occupied is special: the statistical factors for excitonic and green0 emissions are negligible at all times
(they are on the order of 0.5$\%$) and only green1 light is emitted periodically, with $P_{g1}(t)$ following $n_{CB}(t)$. We have found this for all values of $U_{eff}$,
either negative or positive, as illustrated in Fig. \ref{fig8}. This behavior can be understood by looking at the electron occupancies. For negative values of $U_{eff}$, 
as the one shown in the left panels of Fig. \ref{fig8}, there are times at which $n_{CB}(t) \simeq 0$, $n_{T}(t) \simeq 1.7$ and $n_{VB}(t) \simeq 1.3$. At these times 
$P_{exc}(t) \simeq 0$ and $P_{g1}(t) \simeq 0$ because there are no electrons in the CB and and $P_{g0}(t) \simeq 0$ because the trap is nearly double-occupied, 
this state not being accessible for emission of green0 light, and, moreover, the VB contains more than one electron.
At times when $n_{CB}(t)$ has a maximum, $n_{VB} \simeq 2$ thus preventing exciton and green0.  Hence the emission of green1 
light is the only radiative deexcitation process at any time, in spite of the correlated electron dynamics. 
Actually, from our calculations we know the weight of each of the states in the time-dependent wavefunction Eq. (\ref{psi3e-t}) and find that only the states 
DD0, D00T and D0TT have an appreciable weight. The states DD0 and D00T have two electrons in the VB and, therefore, are not accessible as final states 
for exciton or green0 emissions. The state D0TT has two electrons in the trap and is not accessible for light emission of any energy. 
However, the state DD0 has an electron in the CB and no electrons in the trap and, consequently, is accessible for emission of green1 light. $P_{g1}(t)$ depends on time because the system fluctuates between the three states. 
For positive values of $U_{eff}$ (right panels in Fig. \ref{fig8}), $n_{VB} \simeq 2$ $\forall t$ 
(actually the system dynamics for $U_{eff} \gtrsim 0$ is basically that of a single electron hopping back and forth between the CB and the trap) 
and this fact leaves green1 light as the only possible emission at any time. The emission is periodic with on/off cicles and a period that depends on the system parameters. Thus, the light emission statistical factors obey a "single-electron motion" picture of one electron hopping between CB and trap, irrespective of the actual correlated dynamics of the three-electron system shown for negative values of $U_{eff}$.
The absence of an excitonic line is independent of the hopping $V_{CB}$,
as illustrated in Fig. \ref{figA3} in Appendix D, where we have increased this parameter by a factor of 2. Also, note in Fig. \ref{figA3} that the values of the time-averaged occupancies have nor changed much by this change in the hopping.
At this point we should mention that additional calculations using as initial state the other state having a double-occupied VB, state $D00T,n=0$, produce exactly the same results, namely only green1 light is emitted periodically for any value of $U_{eff}$.
Finally Figure \ref{figA4} in Appendix E compares calculations performed with two values of the hopping parameters differing by a factor of two, showing that the transient time increases by the same factor when we decrease the hopping while the stationary regime is largely unaffected.

In the light of our results, we attempt to give a possible explanation for the increase in the intensity of exciton emission with the concomitant decrease in the intensity of the green emission lines upon continuous illumination with ultraviolet radiation, 
experimentally observed for ZnO nanoparticles suspended in an alcohol. At the beginning of the illumination an electron is photoexcited to the CB leaving a hole in the VB.
This hole can be half filled at most, in times on the order of tens of femtoseconds, as a consequence of electron-phonon interaction.
 The rest of the hole is then quenched by electrons from the alcohol in times
$\lesssim 15$ ps  \cite{JACS_Schimpf}. In this way, the much slower radiative processes take place with a fully occupied VB and, consequently, only green1 light can be emitted.
By continuous illumination of the sample more and more holes are produced in the VB yielding situations
that could be modeled by other of the initial 
conditions investigated in this work, all of them giving a statistical factor for excitonic emission larger than the ones for the green emissions. 
In this way an excitonic line could show up with an intensity larger than the green emission one. 
However, the actual intensities of the emission lines are proportional to the matrix elements 
$|\langle \phi_r|\hat h_{rad}|\phi_s\rangle|^2 $, where $|\phi_r\rangle$ and $|\phi_s\rangle$ are single-particle states (see Eq. (15)), not considered in this work, and thus 
a strong modification of the single-particle wavefunctions due to the accumulation of photoexcited electrons in the CB, 
will be of paramount importance.

\section{Conclusions}
\label{sec-conclusions}

In this work we have analyzed the effects of electron-electron and electron-phonon interactions in the dynamics of a system of two or three electrons that can be trapped to a localized state and detrapped to extended band states using a simple model, in which the QD's valence and conduction bands are  modeled by two single-particle energy levels and the trap is described by one single-particle level within the band gap. 
In spite of its simplicity the time dependent model has no analytical solution but a numerically exact one can be found at a relatively low computational cost. Within this model, we study the time evolution of the electron occupancies of the CB, VB and trap, as well as the statistical factors influencing the emission of light of different energies in this system. 
Electron-phonon interaction at the trap introduces electronic sub-levels near the valence and conduction bands that facilitates an efficient transfer of electrons among the electronic levels and possibilitates that an initially empty VB gets up to 0.7 electrons.
In most of the analyzed  cases, the system dynamics has a very short transient determined by the hopping parameters, that can be of tens of femtoseconds,
followed by a quasi-stationary regime in which the electron occupancies either oscillate periodically around their time-averaged values or remain nearly constant. 
The quasi-stationary values depend on the initial state of the electronic system and on other system parameters as well but are generally rather independent of the hoppings. 
We find signatures of strong electronic correlations in the electronic motion for negative values of the effective electron-electron Coulomb interaction that are not translated to the statistical factors for light emission. These factors always show fast oscillations associated to electrons hopping back and forth from the CB to the trap, irrespective of the usually slower motion of the VB electrons. 
Our calculations show that light emission of both exciton and green lines are possible in all the investigated cases except in the especial cases in which the VB is initially filled with two electrons. In these cases the VB can lose and recover electrons periodically but exciton emission is negligible at any time. We use this fact to attempt to give a possible explanation for the increase in the intensity of exciton emission with the concomitant decrease in the intensity of the green emission lines upon continuous illumination with ultraviolet radiation, 
experimentally observed for ZnO nanoparticles suspended in an alcohol.

\section*{Acknowledgments}
 Financial support from the Spanish
Ministry of Science and Innovation through the Mar\'ia de Maeztu Program for Units of Excellence in R$\&$D
(CEX2023-001316-M) and the project PID2024-156077OB-I00 (DQUOTE) is acknowledged. 

\section*{Data Availability}
The data that support the findings of this article are openly
available in the arXiv repository [51]

\newpage

\section*{Appendixes}
\subsection{}
This is the system of coupled linear differential equations we have to solve for calculating the electron-phonon dynamics in the case of three electrons.

\begin{eqnarray}
\frac{d a_{DD0,n}(t)}{dt}&=&-i(2\epsilon_{VB}+\epsilon_{CB}+n\omega_0)a_{DD0,n}(t)-i V_{CB} a_{D00T,n}(t)-i V_{VB}[a_{b,n}(t)+a_{c,n}(t)], \nonumber \\
\frac{d a_{DD1,n}(t)}{dt}&=&-i(\epsilon_{VB}+2\epsilon_{CB}+n\omega_0)a_{DD1,n}(t)-i V_{VB} a_{D11T,n}(t)-i V_{CB} [a_{a,n}(t)+a_{c,n}(t)],  \nonumber \\
\frac{d a_{D00T,n}(t)}{dt}&=&-i(2\epsilon_{VB}+\epsilon_T+n\omega_0)a_{D00T,n}(t)-i V_{VB} a_{D0TT,n}(t)-i V_{CB}a_{DD0,n}(t) \nonumber \\
                         & &-i \lambda \sqrt{n}a_{D00T,n-1}(t)-i \lambda \sqrt{n+1}a_{D00T,n+1}(t), \nonumber \\
\frac{d a_{D11T,n}(t)}{dt}&=&-i(2\epsilon_{CB}+\epsilon_T+n\omega_0)a_{D11T,n}(t)-i V_{CB} a_{D1TT,n}(t)-i V_{VB} a_{DD1,n}(t)] \nonumber \\
                         & &-i \lambda \sqrt{n}a_{D11T,n-1}(t)-i \lambda \sqrt{n+1}a_{D11T,n+1}(t), \nonumber \\
\frac{d a_{a,n}(t)}{dt}&=&-i(\epsilon_{VB}+\epsilon_{CB}+\epsilon_T+n\omega_0)a_{a,n}(t)-i  V_{VB} a_{D0TT,n}(t)-i V_{CB} a_{DD1,n}(t) \nonumber \\
                         & &-i \lambda \sqrt{n}a_{a,n-1}(t)-i \lambda \sqrt{n+1}a_{a,n+1}(t), \nonumber \\
\frac{d a_{b,n}(t)}{dt}&=&-i(\epsilon_{VB}+\epsilon_{CB}+\epsilon_T+n\omega_0)a_{b,n}(t)-i  V_{CB} a_{D1TT,n}(t)-i V_{VB} a_{DD0,n}(t) \nonumber \\
                         & &-i \lambda \sqrt{n}a_{b,n-1}(t)-i \lambda \sqrt{n+1}a_{b,n+1}(t), \nonumber \\ 
\frac{d a_{c,n}(t)}{dt}&=&-i(\epsilon_{VB}+\epsilon_{CB}+\epsilon_T+n\omega_0)a_{c,n}(t)-i  V_{VB} [a_{D1TT,n}(t)+a_{DD0,n}] \nonumber \\
                         & &-i V_{CB} [a_{D0TT,n}(t)+a_{DD1,n}(t)] \nonumber \\
                         & &-i \lambda \sqrt{n}a_{c,n-1}(t)-i \lambda \sqrt{n+1}a_{c,n+1}(t), \nonumber \\                        
\frac{d a_{D0TT,n}(t)}{dt}&=&-i(\epsilon_{VB}+2 \epsilon_T+U+n\omega_0)a_{D0TT,n}(t)-i V_{VB} [a_{D00T,n}(t)+ a_{a,n}(t)] -i V_{CB} a_{c,n}(t) \nonumber \\
                         & &-i 2 \lambda \sqrt{n}a_{D0TT,n-1}(t)-i 2 \lambda \sqrt{n+1}a_{D0TT,n+1}(t), \nonumber \\                                               
\frac{d a_{D1TT,n}(t)}{dt}&=&-i(\epsilon_{CB}+2 \epsilon_T+U+n\omega_0)a_{D1TT,n}(t)-i V_{CB} [a_{D11T,n}(t)+a_{b,n}(t)] -iV_{VB} a_{c,n}(t) \nonumber \\
                         & &-i 2 \lambda \sqrt{n}a_{D1TT,n-1}(t)-i 2 \lambda \sqrt{n+1}a_{D1TT,n+1}(t), \nonumber \\                                               
\label{A1}
\end{eqnarray}

\subsection{}
\begin{figure}[htbp]
\centering
\includegraphics[width=80mm]{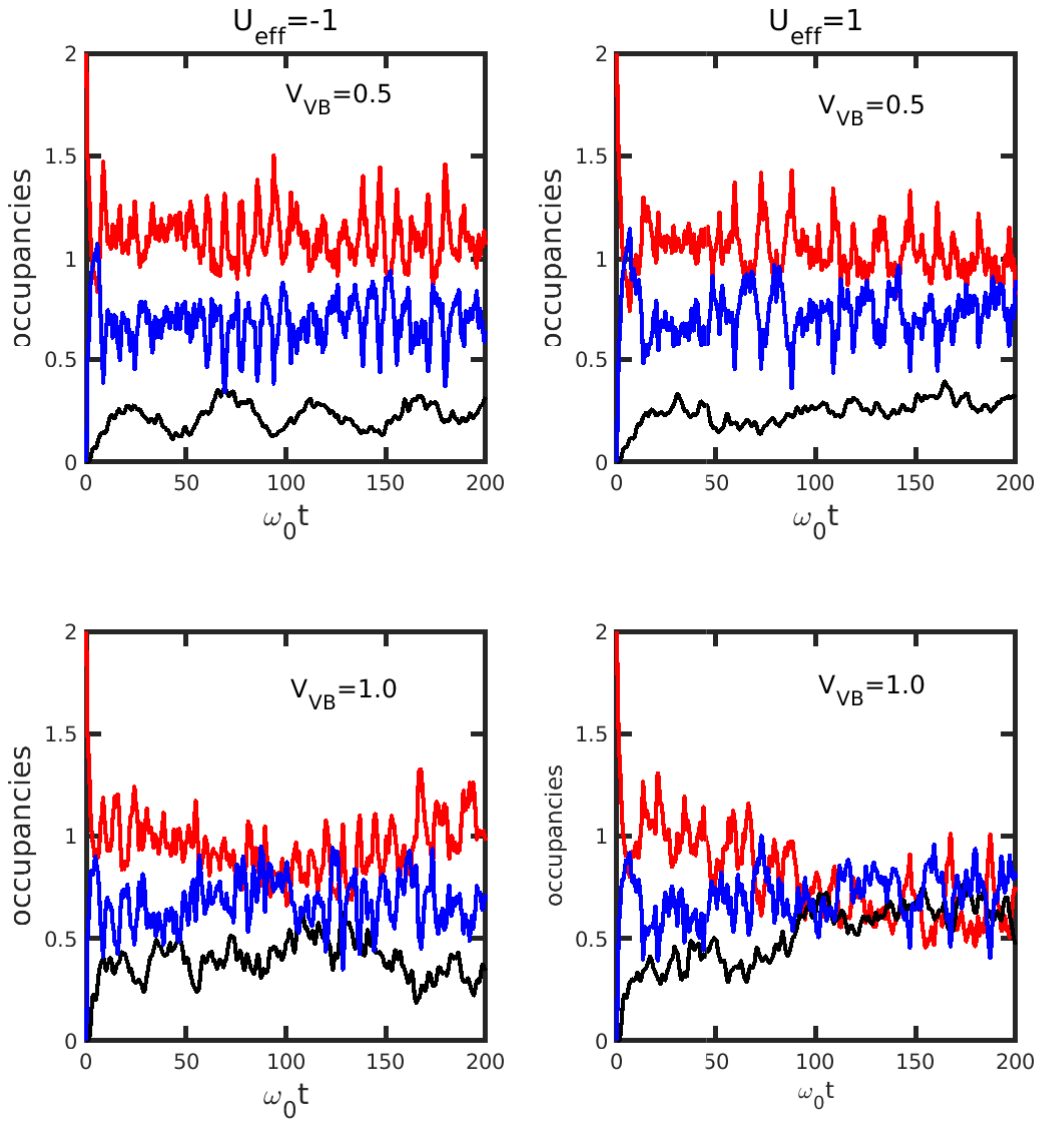}
\caption{\label{figA1} Time dependent occupancies $n_{CB}(t)$ (red lines), $n_{VB}(t)$ (black lines) and $n_{T}(t)$ (blue lines)
for four cases all with $\tilde \epsilon_{T}=10 \omega_0$, $E_{g}= 20 \omega_0$, $V_{CB}=\omega_0$  and $\lambda=3 \omega_0$,
differing in $V_{VB}$ and in $U_{eff}$. 
The initial state has two electrons in the CB ($DD2,n=0$ in Fig. \ref{esq1}). 
The two top panels are for $V_{VB}=0.5 \omega_0$ and two bottom panels are for $V_{VB}= \omega_0$.
The two left panels are for $U_{eff}= -1 \omega_0$ and the two right panels are for $U_{eff}= 1 \omega_0$
.} 
\end{figure}

Figure \ref{figA1} shows the system dynamics of two electrons in cases where $|U_{eff}| \simeq V_{VB}$. The initial state has two electrons in the CB.
The correlated/uncorrelated dynamics depends on 
the sign of $U_{eff}$ and also on $V_{VB}$.

\subsection{}
\begin{figure}[htbp]
\centering
\includegraphics[width=80mm]{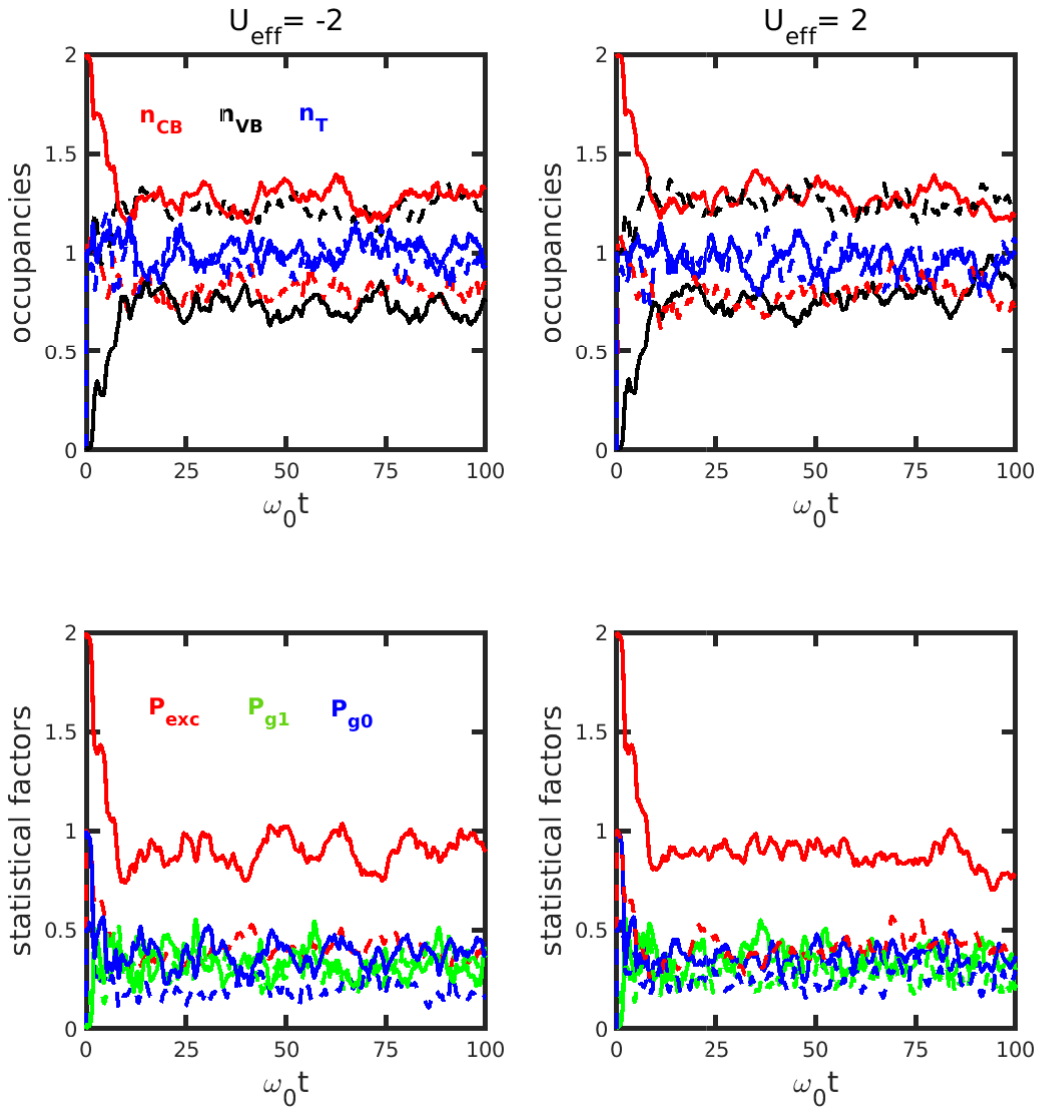}
\caption{\label{figA2} Time dependent occupancies (top panels) and light emission statistical factors (bottom panels) for a three electron system with
$\tilde \epsilon_{T}=10 \omega_0$, $E_{g}= 20 \omega_0$, $V_{VB}= V_{CB}=\omega_0$, $\lambda=3 \omega_0$, two values of $U_{eff}$ 
and two initial conditions drawn as:
dashed lines are for an electron in the trap, another in the CB and the other in the VB (state $a,n=0$ in Fig. \ref{esq3}); 
continuous lines are for initial state with two electrons in the CB and the other in the trap (state $D11T,n=0$ in Fig. \ref{esq3}).
Left panels are for $U_{eff}=-2 \omega_0$ and right panels are for $U_{eff}=2 \omega_0$.
In the top panels red lines plot $n_{CB}$, blue lines plot $n_{T}$ and black lines plot $n_{VB}$.
In the bottom panels red lines plot $P_{exc}$, green lines plot $P_{g1}$ and blue lines plot $P_{g0}$
.} 
\end{figure}

Figure \ref{figA2} shows the dynamics of a three electron system with two different initial conditions. In both cases we find
a quasi-stationary dynamics, with all the three occupancies and the three statistical light emission factors 
reaching very fast nearly constant values, in times $t \sim \frac{10}{V_{CB}}$, 
this kind of dynamics being insensitive to the sign of $U_{eff}$.  

\subsection{}
\begin{figure}[htbp]
\centering
\includegraphics[width=80mm]{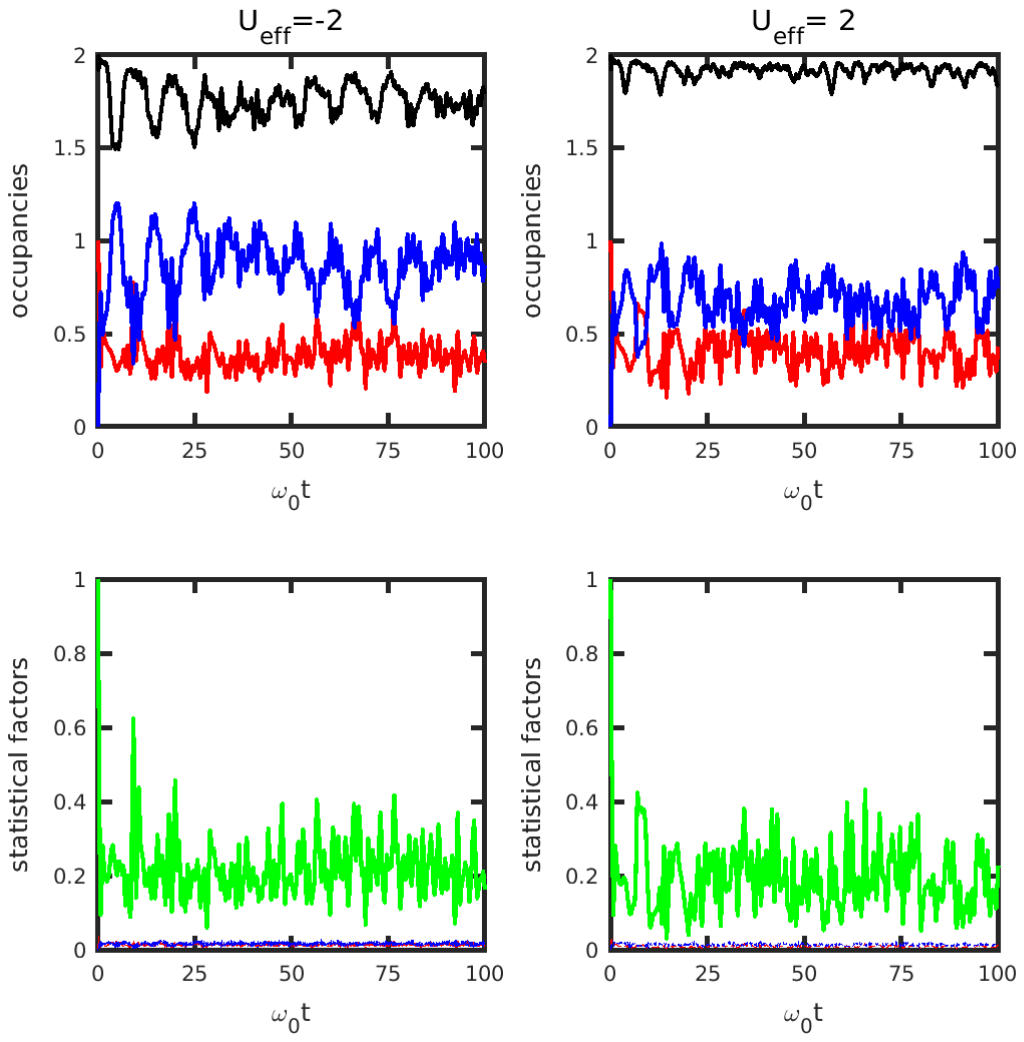}
\caption{\label{figA3} Time dependent occupancies (top panels) and light emission statistical factors (bottom panels) for a three electron system with
$\tilde \epsilon_{T}=10 \omega_0$, $E_{g}= 20 \omega_0$, $V_{VB}= V_{CB}=2 \omega_0$, $\lambda=3 \omega_0$, two values of $U_{eff}$. 
The initial state has two electrons in the VB and the other in the CB (state $DD0,n=0$ in Fig. \ref{esq3}).
Left panels are for $U_{eff}=-2 \omega_0$ and right panels are for $U_{eff}=2 \omega_0$.
In the top panels red lines plot $n_{CB}$, blue lines plot $n_{T}$ and black lines plot $n_{VB}$.
In the bottom panels red lines plot $P_{exc}$, green lines plot $P_{g1}$ and blue lines plot $P_{g0}$
.} 
\end{figure}

Figure \ref{figA3} shows the dynamics of a three electron system with an initially double-occupied VB (state $DD0,n=0$ in Fig. \ref{esq3})
for the same parameters as in Fig. \ref{fig8} except for the hoppings that are a factor of two larger.
Note that increasing the hoppings does not affect the time-averaged values of electron occupancies, that are very similar to the ones in Fig. \ref{fig8}.
Since in this case $|U_{eff}|=V_{VB}=V_{CB}$ the electron-phonon dynamics is less correlated than in Fig. \ref{fig8} and only green1 light can be emitted. 

\subsection{}
\begin{figure}[htbp]
\centering
\includegraphics[width=80mm]{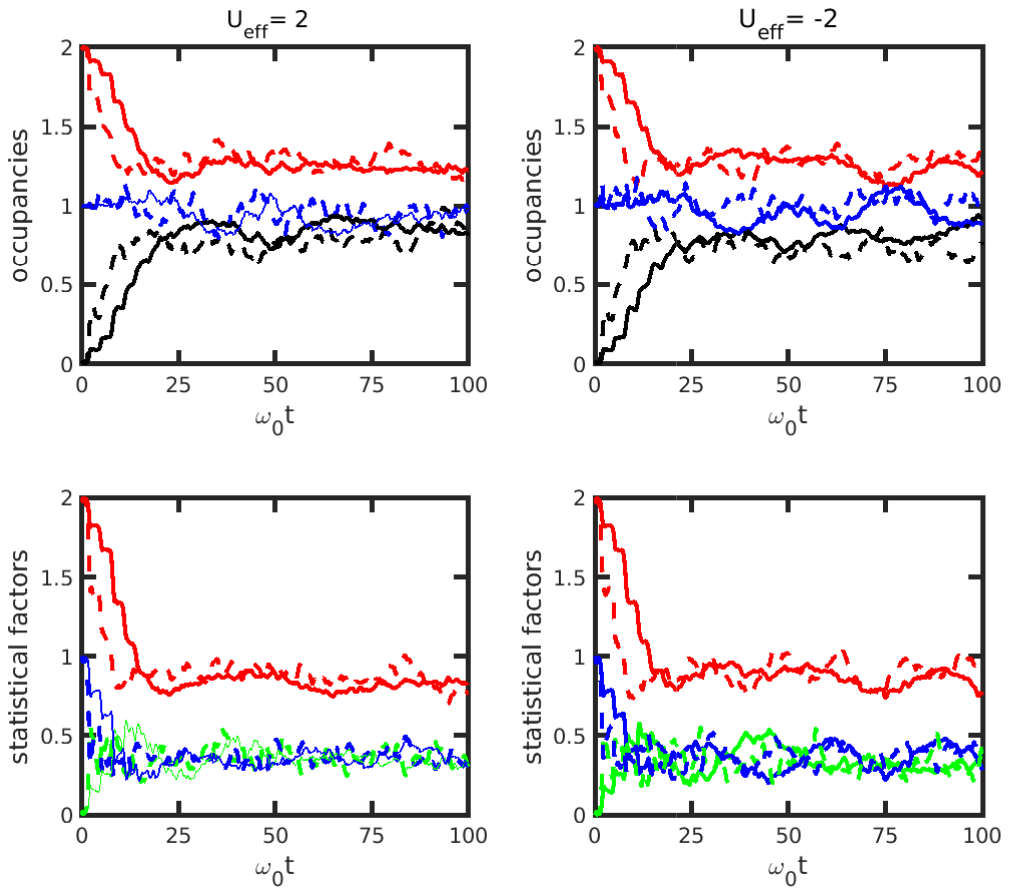}
\caption{\label{figA4} Time dependent occupancies (top panels) and light emission statistical factors (bottom panels) for a three electron system with
$\tilde \epsilon_{T}=10 \omega_0$, $E_{g}= 20 \omega_0$, $\lambda=3 \omega_0$ and $U_{eff}=\pm 2\omega_0$. 
The initial state has two electrons in the CB and the other in the trap (state $D11T,n=0$ in Fig. \ref{esq3}).
Left panels are for $U_{eff}=2 \omega_0$ and right panels are for $U_{eff}=-2 \omega_0$. Continuous lines are for 
$V_{CB}=V_{VB}= 0.5 \omega_0$ and dotted lines are for $V_{CB}=V_{VB}=\omega_0$. 
In the top panels red lines plot $n_{CB}$, blue lines plot $n_{T}$ and black lines plot $n_{VB}$.
In the bottom panels red lines plot $P_{exc}$, green lines plot $P_{g1}$ and blue lines plot $P_{g0}$.
Deacreasing the hopping parameters by a factor of two inceases the transient time by the same factor. 
However, occupancies and light emission factors are not affected in the quasi-stationary regime 
.} 
\end{figure}

Figure \ref{figA4} compares the dynamics of a three electron system for two values of $V_{CB}=V_{VB}$ differing by a factor of two.
 Continuous lines are for 
$V_{CB}=V_{VB}= 0.5 \omega_0$ and dotted lines are for $V_{CB}=V_{VB}=\omega_0$.
The initial state has two electrons in the CB and one in the trap (state $D11T,n=0$).
Decreasing the hopping parameters by a factor of two increases the transient time by the same factor. However, the occupancies and the light emission factors are not affected in the quasi-stationary regime.

 \newpage


\end{document}